\documentclass[aps,amsfonts,amsmath,prd,preprint,nofootinbib]{revtex4}
\newcommand{\beq}{\begin{equation}}
\newcommand{\eeq}{\end{equation}}

\usepackage{epsfig,bbm}
\begin{document}

\title{Decay of flux vacua to nothing}
\author{\footnote[0]{Electronic addresses:}Jose J.  Blanco-Pillado\footnote[0]{\tt jose@cosmos.phy.tufts.edu}, Handhika S. Ramadhan\footnote[0]{\tt handhika.ramadhan@tufts.edu}, and Benjamin Shlaer\footnote[0]{\tt shlaer@cosmos.phy.tufts.edu}}
\affiliation{Institute of Cosmology, Department of Physics and Astronomy,\\ 
Tufts University, Medford, MA 02155, USA}

\def\changenote#1{\footnote{\bf #1}}

\begin{abstract}
We construct instanton solutions describing the decay of 
flux compactifications of a $6d$ gauge theory by generalizing the Kaluza-Klein
bubble of nothing. The surface of the bubble is described by 
a smooth magnetically charged solitonic brane whose asymptotic
flux is precisely that responsible for stabilizing the $4d$ 
compactification. We describe several instances of
bubble geometries for the various vacua occurring in a 
$6d$ Einstein-Maxwell theory namely, $AdS_4 \times S^2$, 
${\mathbb R}^{1,3} \times S^2$, and $dS_4 \times S^2$. Unlike 
conventional solutions, the bubbles of nothing introduced here 
occur where a {\em two}-sphere compactification manifold homogeneously
degenerates.

\end{abstract}

\maketitle
\thispagestyle{empty}
\section{Introduction}
\setcounter{page}{1}

The necessity of extra dimensions has strong theoretical backing in the
context of string theory, but stabilizing the shape and size moduli of
the compactification manifold has historically been one of the most
challenging obstacles for realistic model building.
Field theories with higher rank fluxes wound on internal cycles were proposed long 
ago as a remedy to this problem \cite{Cremmer, Freund, ME-6d, GellMann}.
Similar mechanisms have been incorporated in string theory
compactifications \cite{GKP, KKLT, Douglas} which suggest 
the existence of a tremendous multitude of stable and metastable vacua, the 
so-called string landscape \cite{Landscape}. The interplay between 
these solutions and eternal inflation \cite{eternal-inflation-alex,eternal-inflation-andrei} 
opens the possibility for transitions between the various flux-vacua
\cite{Kachru, Frey, BPSPV-1,IS-Yang, Dahlen-Brown}. Furthermore, it has been
found that there exist more exotic classes of transitions which 
change the effective dimensionality of spacetime
\cite{Linde-Zelnikov, Giddings, BPSPV-1,CJR, BPSPV-2}. Although work
on these transitions is in its early stages, already it appears they
may have interesting theoretical \cite{SPV} and observational
\cite{GHR,BPMS,Adamek,Salem} consequences.

In this paper we generalize a new decay channel that has 
recently been shown to exist in axionic flux compactifications \cite{BPBS}. 
This instability renders vacua susceptible to decay via the nucleation 
of a generalized bubble of nothing \cite{Witten}, one that is charged 
with respect to the flux which induces the spontaneous
compactification (See also \cite{HOP,IS-Yang} for a discussion of related ideas).

This paper is organized as follows. In section II
we discuss the $6d$ Einstein-Maxwell landscape. In section III
we embed the Maxwell theory in the simplest non-abelian gauge theory, yielding
the Einstein-Yang-Mills-Higgs model of $SU(2)$. In section IV
we describe new instanton configurations in detail and provide explicit numerical
examples within a family of solutions. Finally,
we conclude in section V.

\section{The Einstein-Maxwell landscape in $6d$}

The Einstein-Maxwell theory in $6d$ \cite{ME-6d} is a remarkably simple model which
nevertheless enjoys many important features of more realistic
flux compactifications of string
theory \cite{Douglas}. The action is given by

\beq
S =\int{d^6  x \sqrt{-g} \left(   {1\over {2\kappa^2}} 
  R - {1\over 4} F_{MN} F^{MN} -  \Lambda\right)}\,,
\label{EM-6D-action}
\eeq
where our conventions are as follows.  Six dimensional indices are
indicated with capital latin letters, $M,N = 0...5$.  The $6d$ reduced 
Planck mass is written $M_{(6)} = 1/\sqrt{\kappa}\,\,$, and $\Lambda$
is the six dimensional cosmological constant, which we will assume to 
be non-negative. 

This model was explored in detail in \cite{BPSPV-1, BPSPV-2,SPV},
where it was shown to possess distinct families of flux
compactifications: a magnetic sector with geometry
$(A)dS_4 \times S^2$ or ${\mathbb R}^{1,3} \times S^2$, an electric sector with
spacetime $AdS_2 \times S^4$, and a higher dimensional vacuum with
no flux, $dS_6$. Several possible transitions between these sectors were discussed in
\cite{BPSPV-1,CJR, BPSPV-2,IS-Yang, Dahlen-Brown}, which suggest the existence of a 
complex multi-dimensional landscape even in this simple model. 

Here we study a new decay channel for the $4d$ flux vacua, 
the nucleation of a  bubble of nothing \cite{Witten}. The portion of 
this landscape under consideration is the magnetic
sector (four large dimensions), which we will now review.

The equations of motion obtained from the action in Eq.~(\ref{EM-6D-action}) are
\begin{eqnarray}
R_{MN} - {1\over 2} g_{MN} R &=& \kappa^2  T_{MN}\,, \\
{1\over{\sqrt{- g}}} \partial_M \left(\sqrt{-g} F^{MN}\right) &=&0\,\,,
\end{eqnarray}
with energy-momentum tensor
\beq
T_{MN} ={g}^{LP} F_{ML} F_{NP} - {1\over 4} {g}_{MN} F^2 
- {g}_{MN} {\Lambda}~.
\eeq 
In the magnetic sector, the metric takes the form
\beq
ds^2= {g}_{MN} dx^M dx^N = {g}_{\mu \nu} d x^{\mu}
d x^{\nu} + C^2 d\Omega_2^2~,
\label{6D-metric}
\eeq 
where $g_{\mu \nu}$ describes a four dimensional maximally symmetric space,\hspace{-1mm}
\footnote{The $4d$ part of the metric has Ricci scalar 
$R^{(4)} =  12  H^2$, where $H^2$  may be negative.}
and the compactification manifold is a 2-sphere of radius $C$.

The field strength in this sector is given by the {\it
  monopole}-type configuration \cite{ME-6d},
\beq
F_{\theta \phi} = -F_{\phi \theta} = {n \over{2 e}} \sin{\theta}\,\,,
\eeq
which respects the chosen isometries of the metric,
and saturates the Dirac quantization condition $\int_{S^2}F = 2\pi
n/e$, where $n\in \mathbb{Z}$ and $e$ is the quantum of
electric charge. With this ansatz, the electromagnetic equations 
of motion are automatically satisfied, and the Einstein equations 
lead to the relations for $H$ and $C$
\begin{eqnarray}
3 H^2 + {1\over {C^2}} = \kappa^2 \left({{n^2}\over{8 e^2 C^4}} +
 \Lambda\right)\,, \\
6 H^2  =  \kappa^2 \left(\Lambda - {{n^2}\over{8 e^2 C^4}}\right)~.
\end{eqnarray}
This can be solved in terms of the parameters of the 
$6d$ theory and the magnetic flux number $n$, yielding the 
solutions
\begin{eqnarray}
C^2 &=& {1\over{ \kappa^2 \Lambda}} \left(1 \mp \sqrt{1 - {{3 n^2}
\over {4 n_0^2}}}\right)\,,\nonumber\\
H^2 &=& {{2 \kappa^2 \Lambda}\over {9 }} \left[1 -  {{2 n_0^2 }\over {3
 n^2}} \left(1 \pm \sqrt{1 -  {{3 n^2 }\over {4
        n_0^2}}}\right)\right]~,
\label{EM-landscape-solutions}
\end{eqnarray}
where we have defined
\beq
n_0^2 = {{2 e^2 }\over {\kappa^4 \Lambda}}~.
\label{n0}
\eeq

\begin{figure}[htbp]
\centering\leavevmode
\epsfysize=8cm \epsfbox{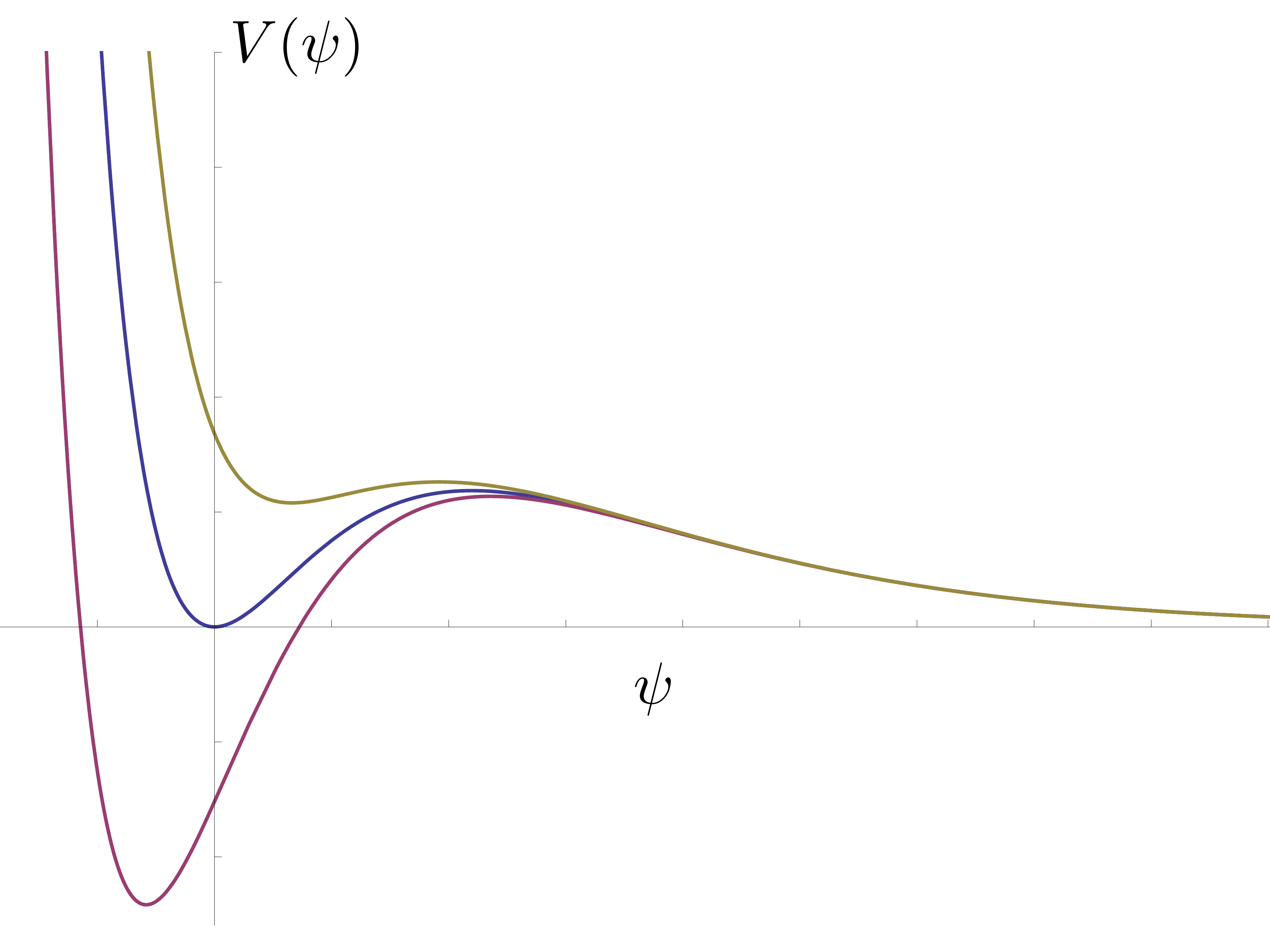}
\caption {Plot of the $4d$ effective potential in reduced Planck units,
as a function of the field $\psi$ for three different values of monopole
number $n$.}
\label{6D-potential}
\end{figure}

The twofold existence of solutions when $\Lambda > 0$ can be
understood by looking at Fig.~\ref{6D-potential}, the $4d$ effective potential
for the radion, which governs the size of the extra dimensions.
Following \cite{BPSPV-1}, we generalize the six dimensional metric ansatz to
\beq
ds^2= g_{MN} dx^M dx^N = e^{- \psi(x)/M_P} g^{(4)}_{\mu \nu}
dx^{\mu} dx^{\nu} + e^{\psi(x)/M_P} C_0^{2}~d\Omega_2^2\,\,,
\eeq
with
$C_0 =1 /\sqrt{2 \kappa^2 \Lambda}~$. Together 
with the monopole-type configuration for the Maxwell field, this ansatz allows 
integration of the full $6d$ action 
over the internal manifold, yielding a $4d$ effective theory
with low energy action
\beq
S= \int{d^4 x \sqrt{-g^{(4)}}\left({1\over 2} M_P^2 R^{(4)} - {1\over 2}
  \partial_{\mu} \psi \partial^{\mu} \psi - V(\psi)\right)}~,
\eeq
with the potential for the canonical radion $\psi$ given by
\beq
V(\psi)= {{4\pi}\over {\kappa^2}} \left({{n^2}\over{2 n_0^2}}
e^{-3\psi/M_p} - e^{-2\psi/M_P} + {1 \over 2} e^{-\psi/M_P} \right)\,.
\eeq
The $4d$ Planck mass $M_P$ is dependent on
the volume of the compactification manifold 
via $M_P^2 = 4 \pi C^2 /\kappa^2$.  
In Fig.~(\ref{6D-potential}), we plot the effective potential $V(\psi)$ for 
three choices of flux number $n$.
We can immediately see that at most one of the two
solutions shown in Eqs.~(\ref{EM-landscape-solutions}) can be stable, while the
other, once perturbed, will roll to either the stable solution
or decompactification.  Henceforth, we consider only the stable solutions to the
equations of motion.\footnote{Perturbative stability of flux
compactifications has been discussed previously in \cite{perturbative-stability}.}

Transitions between flux vacua are mediated by instantons
constructed \cite{BPSPV-1} from the magnetically 
charged black 2-branes known to exist in the spectrum of the 
theory \cite{Gibbons,Gregory}.  On the other hand, it was recently
suggested in \cite{BPBS} that there should be a special transition that
would decrease the flux number of the compactification to zero. It
is clear that if such a transition occurs, there will be no obstacle
for the internal geometry to collapse and create a
large coordinate region of volume measure zero, a bubble of nothing
\cite{Witten}. Furthermore, the surface of this bubble must act as a source 
for the magnetic flux present in the asymptotic region of the
compactification, and so we generalize the bubble of 
nothing to include this charge.  Recently, two of us have demonstrated the existence
of charged bubbles of nothing in a simple axionic flux
compactification. In that fully backreacting $5d$ solution, the
surface of the bubble is a de Sitter vortex charged with respect 
to the axion \cite{BPBS}. In this paper, we generalize the instability 
to the more realistic landscape of the  $6d$ Einstein-Maxwell theory. 
One clear candidate for the bubble of nothing is a
generalization of the well-known codimension three Dirac monopole.
We require a solution where the extra-dimensional spacetime
is smooth everywhere, in particular in the region where the 2-sphere 
degenerates to zero size. This is difficult to achieve in the Einstein-Maxwell model,
since it seems to inevitably lead to a singularity at the location of the
monopole. We solve this problem in a natural way by introducing new degrees of freedom
which resolve the singularity: by embedding the model in a non-abelian 
gauge theory which is known to possess smooth magnetically charged
solitons of codimension three, the Yang-Mills-Higgs model
\cite{'tHooft, Polyakov}, which we will now review.

\section{The Einstein-Yang-Mills-Higgs Landscape} 

One can imagine an embedding of the Einstein-Maxwell theory
presented in the previous section into more complicated models which
include new degrees of freedom only in the UV, and so would not
distort the landscape of $4d$ flux vacua computed previously. Here we realize
this with a specific Einstein-Yang-Mills $SU(2)$ model with an adjoint Higgs breaking the
gauge symmetry to $U(1)$, which we identify with the Maxwell field described
above. This is one of the first flux compactification models described in the 
literature \cite{Cremmer}, and as we will see, it is well suited to our goal
of finding a UV completion of the Einstein-Maxwell flux vacuum instability known
as a bubble of nothing.

The model is defined by the action
\begin{equation}
{S}=\int d^{6}x\sqrt{-g}\left(\frac{1}{2\kappa^2}R
-\frac{1}{4}{\cal F}^{a}_{MN}{\cal F}^{a
  MN}-\frac{1}{2}D_{M}\Phi^{a}D^{M}\Phi^{a} - V(\Phi) - \Lambda\right),
\end{equation}
with
\begin{eqnarray}
V(\Phi)& = &\frac{\lambda}{4}\left(\Phi^{a}\Phi^{a}-\eta^{2}\right)^{2}\,,\nonumber\\
{\cal F}^{a}_{MN}& = &\partial_{M}A^{a}_{N}-\partial_{N}A^{a}_{M}+ 
e\epsilon^{abc}A^{b}_{M}A^{c}_{N}\,,\\
D_{M}\Phi^{a}& = &\partial_{M}\Phi^{a}+e\epsilon^{abc}A^{b}_{M}\Phi^{c}~.\nonumber
\end{eqnarray}

Varying the action with respect to the fields yields the equations of motion
\begin{eqnarray}
\label{eq:einstein}
R_{AB}-\frac{1}{2}g_{AB}R&=&\kappa^{2}T_{AB}\,,\\
\label{eq:higgs}
\frac{1}{\sqrt{-g}}D_{M}\left(\sqrt{-g}D^{M}\Phi\right)^{a} &=& \lambda \Phi^{a}
\left(\Phi^{b} \Phi^{b}- \eta^2\right)^2\,,\\
\label{eq:yangmills}
\frac{1}{\sqrt{-g}}D_{N}\left(\sqrt{-g}{\cal
  F}^{MN}\right)^{a}&=&e\epsilon^{abc}\left(D^{M}\Phi^{b}\right)\Phi^{c}~,
\end{eqnarray}
where the energy-momentum tensor is given by
\begin{equation}
T_{AB} = D_{A}\Phi^{a}D_{B}\Phi^{a}+{\cal F}^{a}_{AM}{\cal F}_{B}^{a M} +
g_{AB}{\cal{L}}~,
\end{equation}
with
\begin{equation}
{\cal{L}} =  - \frac{1}{2}D_{A}\Phi^{a}D^{A}\Phi^{a} - \frac{1}{4}{\cal F}^{a}_{MN}{\cal F}^{aMN} 
- V(\Phi) - \Lambda~.
\end{equation}

\subsection{Compactification solutions}

In Cremmer {\it et~al.} \cite{Cremmer} it was shown that the preceeding equations lead to a
spontaneous compactification of the $6d$ spacetime after turning on a
monopole-type flux in the spontaneously broken gauge theory, similar to what was
presented in the abelian case of the previous section. Cremmer {\it et~al.} restricted themselves to the flat $4d$ spacetime ${\mathbb R}^{1,3} \times S^2$,
and although their work discussed only the $n=1$ flux compactification, they
managed to find several types of solutions.\footnote{In Appendix
\ref{appendix-n-equal-1} we discuss in detail some of the peculiar properties of this type of
compactification which are special to $n=1$.} Here we generalize such
compactifications to arbitrary integer $n$ flux vacua by choosing a 
matter field ansatz
\begin{eqnarray}\label{n-hedgehog}
\Phi^{a}&=&\eta\ p_{c} (\sin\theta \cos n \varphi, \sin\theta
\sin n \varphi, \cos\theta)\,,\nonumber\\
A^{a}_{\mu}&=&A^{a}_{r}=0\,,\nonumber\\
A^{a}_{\theta}&=&\frac{1-w_{c}}{e} (\sin n \varphi,-\cos n \varphi,0)\,,\\
A^{a}_{\varphi}&=&\frac{n\left(1-w_{c}\right)}{e} \sin\theta (\cos\theta \cos n
\varphi, \cos\theta \sin n \varphi, -\sin\theta)\,,\nonumber
\end{eqnarray}
with $n \in {\mathbbm{Z}}$.
The suitability of this ansatz can be motivated by computing the 
topological charge for this configuration \cite{Arafune} via

\begin{equation}
\frac{1}{4 \pi} \int{d\theta d\phi |\Phi|^{-3} \epsilon_{abc} 
\Phi^a \partial_{\theta} \Phi^b \partial_{\phi} \Phi^c} = n~,
\end{equation}
where $ |\Phi| = \sqrt{\Phi^a\Phi^a}\,$.

Interestingly, for $n>1$ the equations of motion constrain the possible values of
the constants in the ansatz Eq.~(\ref{n-hedgehog}) to be, $p_c = 1$
and $w_c = 0$. The covariant derivative for the scalar triplet then
vanishes, and the energy momentum tensor induced by this
configuration is precisely that found in the abelian flux vacua. This 
can be understood by looking at the form of the 
{\it electromagnetic tensor} \cite{'tHooft}
\begin{equation}
F_{MN} = \frac{\Phi^a}{|\Phi|} {\cal F}^a_{MN} + 
\frac{1}{e |\Phi|^3} \epsilon^{abc} \Phi^a D_M \Phi^b D_N \Phi^c\,,
\end{equation}
which in this case becomes
\begin{equation}
F_{\theta \phi} = {n \over{e}} \sin{\theta}\,,
\end{equation}
and the equations of motion reduce to
\begin{eqnarray}
3 H^2 + {1\over {C^2}} &=& \kappa^2 \left({{n^2}\over{2 e^2 C^4}} +
 \Lambda\right)\,, \\
6 H^2  &=&  \kappa^2 \left(\Lambda - {{n^2}\over{2 e^2 C^4}}\right)~.
\end{eqnarray}

Note that there is a small discrepancy in the definition of the
coupling constant $e$ with respect to the abelian case.
The charge $e$ here is twice the value of the same symbol appearing in the Maxwell theory.
This is reconciled with saturation of Dirac's charge quantization condition by noting that the
smallest-charged particle in the non-abelian theory would be an
$SU(2)$ {\em doublet}, whose charge is equal to $e/2$ using the
present non-abelian convention for $e$. With this dictionary, the
$SU(2)$ theory is indistinguishable from the abelian theory in the IR.

We turn now to discussion of the non-perturbative decay of
flux vacua by the generalized bubble of nothing, and so only consider
the perturbatively stable solutions of the equations of motion. 
Following the arguments presented in the Einstein-Maxwell theory, these are
specified by the two length scales
\begin{eqnarray}
C^2 &=& {1\over{ \kappa^2 \Lambda}} \left(1 - \sqrt{1 - {{3 \kappa^4 \Lambda n^2}
\over {2 e^2}}}\right)\,,\nonumber\\
H^2 &=& {{2 \kappa^2 \Lambda}\over {9 }} \left[1 -  {{ e^2 }\over {3
 \Lambda \kappa^4 n^2}} \left(1 + \sqrt{1 -  {{3 \kappa^4 \Lambda n^2 }\over {2
        e^2}}}\right)\right]~.
\label{EYMH-landscape-solutions}
\end{eqnarray}
The landscape of vacua is identical to the
pure electromagnetic case, in particular we see that the theory has $4d$ compactifications 
$AdS_4 \times S^2$, ${\mathbb R}^{1,3} \times S^2$, and $dS_4 \times S^2$. 

\section{Bubble of Nothing Solutions}

Bubbles of nothing in a simple toy flux compactification were discussed in
\cite{BPBS}, where they were identified as solitonic defects whose intrinsic
worldvolume is a codimension-two de Sitter space. 

We are interested in finding similar objects in a higher dimensional spacetime where
the compactification manifold is a 2-sphere. This leads us to the metric ansatz
\begin{equation}
\label{eq:metric}
ds^{2}= B^{2}(r)(-dt^{2} + \cosh^{2}t~d\Omega_{2}^{2}) + dr^{2}
+ C^{2}(r)(d\theta^{2} + \sin^{2} \theta~d\varphi^{2})~.
\end{equation}

We are searching for solutions that describe the decay of
flux compactifications to a bubble of nothing, i.e., 
solutions where the extra-dimensional space wound with magnetic flux
degenerates to a point at some value of $r$, which we gauge fix to $r = 0$. 
This implies the existence of a magnetic source at the degeneration loci. 
We satisfy this requirement by placing a solitonic magnetic brane
centered at $r = 0$, making use of our UV completion of the low energy
Einstein-Maxwell theory. An appropriate ansatz
in this case is therefore the {\it hedgehog} configuration,
\begin{eqnarray}
\label{eq:matter}
\Phi^{a}&=& \eta\ p(r) (\sin\theta \cos \varphi, \sin\theta
\sin  \varphi, \cos\theta)\,,\nonumber\\
A^{a}_{\mu} &=& A^{a}_{r}=0\,,\nonumber\\
A^{a}_{\theta} &=& \frac{1-w(r)}{e} (\sin  \varphi,-\cos  \varphi,0)\,,\\
A^{a}_{\varphi} &=& \frac{1-w(r)}{e} \sin\theta (\cos\theta \cos \varphi, 
\cos\theta \sin  \varphi, -\sin\theta)~.\nonumber
\end{eqnarray}

For simplicity we are considering only KK-spherically symmetric solutions.  This
demands that we restrict to the case with $n = \pm1$, since higher winding 
solutions are incompatible with spherical symmetry \cite{Weinberg-Guth,Cremmer:1976wp}.
(We do not expect any conceptual difficulty in finding higher $n$ solutions
of reduced symmetry, but they will be more challenging to construct numerically.)

\begin{figure}[htbp]
\centering\leavevmode
\epsfysize=4cm \epsfbox{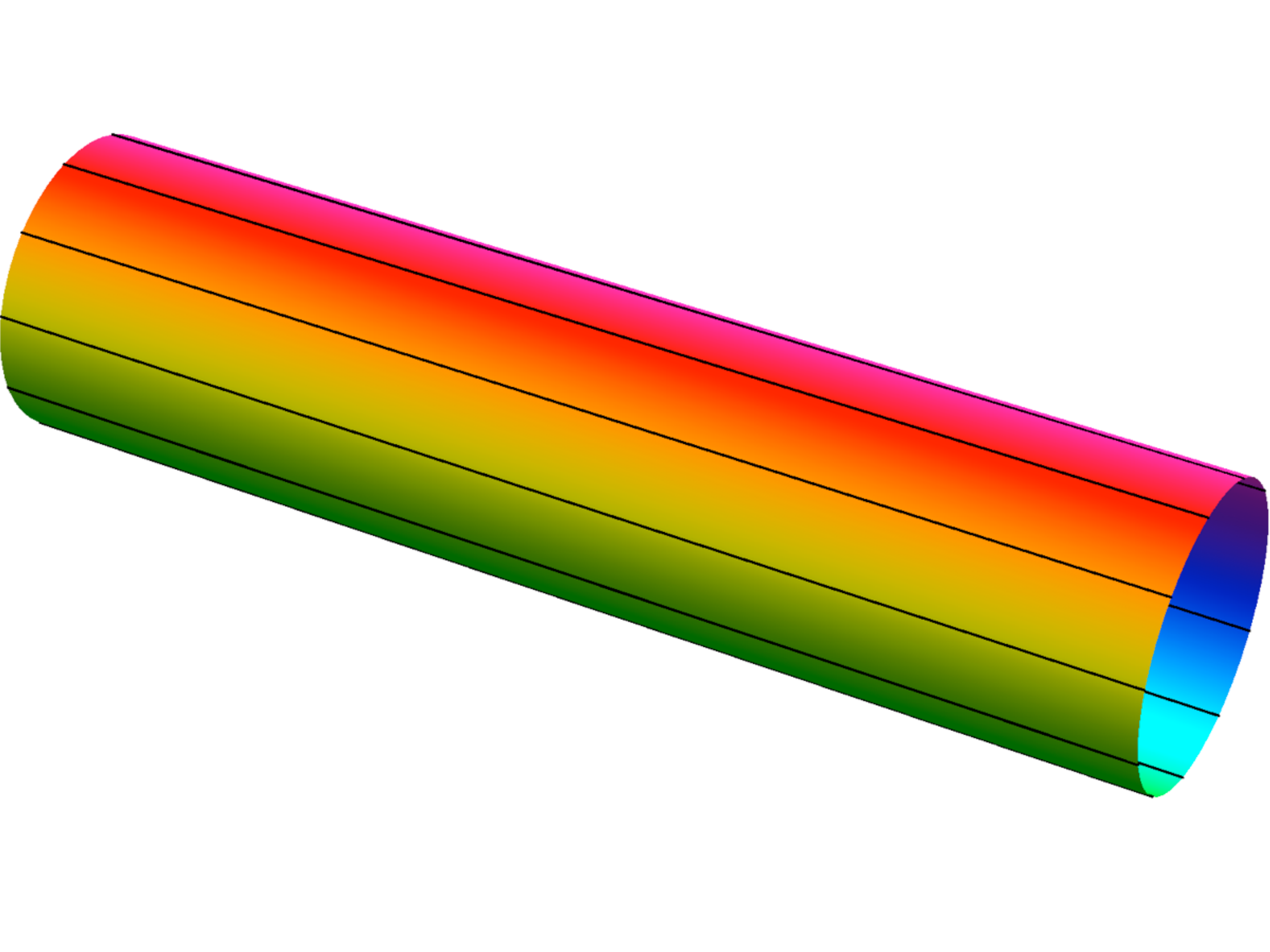}
\hspace{2cm}
\epsfysize=4cm \epsfbox{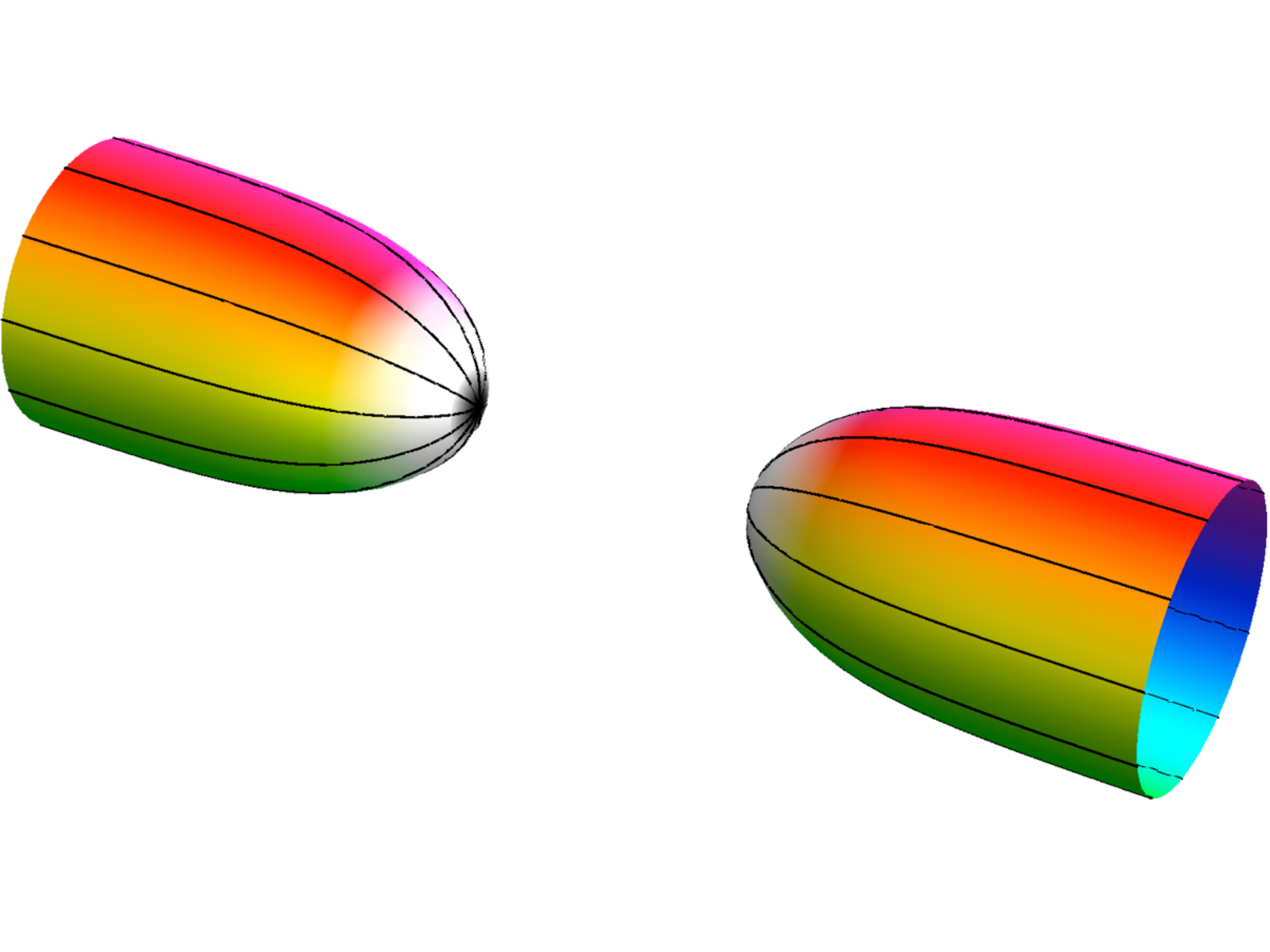}
\caption {Left: An $n=1$ flux compactification, with hue representing
  wound flux.  The $S^2$ compactification manifold is shown as an
  $S^1$, along with one of the four large dimensions.  Right:  A
  bubble of nothing occurs when the compactification manifold
  degenerates. The radion $C(r)$ is plotted as a function of radial
  distance $r$ along the large dimension. Saturation is given by the scalar field magnitude
  $p(r)$, which along with $C(r)$, vanishes at the core of the soliton.}
\label{n1compact}
\end{figure}

Using the above ansatz, the equations of motion for the matter fields
in Eqs.~(\ref{eq:higgs}-\ref{eq:yangmills}) become
\begin{equation}
\label{eq:scalar}
p''+\left(3\frac{B'}{B}+2\frac{C'}{C}\right)p'-\frac{2w^{2}p}{C^{2}}-\lambda\eta^{2}p(p^{2}-1)=0
\end{equation}
and
\begin{equation}
\label{eq:vector}
w''+3\frac{B'}{B}w'+\frac{w (1-w^{2})}{C^{2}}-e^{2}\eta^{2}p^{2}w=0~.
\end{equation}

The Einstein equations are
\begin{eqnarray}
G^{0}_{0} &=& - \frac{1}{B^{2}} - \frac{1}{C^{2}} + \left(\frac{B'}{B}\right)^{2}
+ 4\frac{B'C'}{BC} + \left(\frac{C'}{C}\right)^{2} + 2\frac{B''}{B}+2\frac{C''}{C} 
= \kappa^2 T^{0}_{0}\,,\nonumber\\
G^{r}_{r} &=& - \frac{3}{B^{2}} - \frac{1}{C^{2}} + 3\left(\frac{B'}{B}\right)^{2}
+ 6\frac{B'C'}{BC} + \left(\frac{C'}{C}\right)^{2} = \kappa^2 T^{r}_{r}\, ,\\
G^{\theta}_{\theta} &=& - \frac{3}{B^{2}} + 3\left(\frac{B'}{B}\right)^{2}
+ 3\frac{B'C'}{BC}+3\frac{B''}{B} + \frac{C''}{C}  = \kappa^2 T^{\theta}_{\theta}~,\nonumber
\end{eqnarray}
with energy-momentum tensor specified by
\begin{eqnarray}
T^{0}_{0} &=& - \bigg[\eta^{2}\left(\frac{p'^{2}}{2} + \frac{p^{2}w^{2}}{C^{2}}\right)
+ \frac{1}{e^{2}C^{2}}\left(w'^{2}+\frac{(1-w^{2})^{2}}{2C^{2}}\right) + 
\frac{\lambda \eta^{4}}{4}(p^{2}-1)^{2}+\Lambda\bigg]\,,\nonumber\\
T^{r}_{r} &=& \eta^{2}\left(\frac{p'^{2}}{2} - \frac{p^{2}w^{2}}{C^{2}}\right)
+ \frac{1}{e^{2}C^{2}}\left(w'^{2} - \frac{(1-w^{2})^{2}}{2C^{2}}\right) 
- \frac{\lambda \eta^{4}}{4}(p^{2}-1)^{2}-\Lambda\,,\\
T^{\theta}_{\theta} &=& - \eta^{2}\frac{p'^{2}}{2} + \frac{(1-w^{2})^{2}}{2 e^{2}C^{4}}
- \frac{\lambda \eta^{4}}{4}(p^{2}-1)^{2}-\Lambda~.\nonumber
\end{eqnarray}

Below we will separately study the three different
asymptotic $4d$ effective geometries, $AdS_4,\, {\mathbb R}^{1,3},$ and $dS_4$.
Notice that the asymptotic geometry is specified
once one fixes the values of $\Lambda$ and $e$.  Nevertheless, one may 
find qualitatively different solutions depending on the values of the
other two fundamental parameters, $\eta$ and
$\lambda$. Having explored the form of the solutions in this two
dimensional parameter space, we will comment below on the different
behaviors that one may encounter.

All the solutions we present in this paper have a
magnetically charged soliton at $r=0$, which because of the $2+1$ 
dimensional de Sitter invariance of its world-volume, can be called
{\em inflating}. Inflating braneworld solutions with similar asymptotic behavior
to those presented here have been previously discussed in a 
different context in \cite{alexcho}.

One can show that the most general smooth solution describing the
soliton core has expansion about $r=0$ given by
\begin{eqnarray}\label{nearcore}
p(r) & = & p_{1} r + \cdots\,,\nonumber\\
w(r) & = & 1 + w_{2}r^{2} + \cdots\,,\nonumber\\
B(r) & = & B_{0} + B_2 r^{2}
+ \cdots\,,\\
C(r) & = & r + C_3 r^{3} + \cdots~.\nonumber
\end{eqnarray}
Using the equations of motion we can write all coefficients, 
$B_2$, $C_3$, etc., in terms of three locally undetermined constants, $B_{0}$,
$p_{1}$, and $w_{2}$.
We give the explicit form of these expansions in Appendix 
\ref{appendix-expansion-core}. In the following sections we use the
numerical technique known as the multiple shooting 
method to demonstrate the existence of a solution and determine the values of the
coefficients $B_{0}$, $p_{1}$, and $w_{2}$, such that the matter and
metric fields approach the asymptotic form of the appropriate flux compactification. We
review the multiple shooting method in Appendix \ref{appendix-multiple-shooting}.

\subsection{The decay of $AdS_4 \times S^2$ vacua}

\begin{figure}[htbp]
\centering\leavevmode
\epsfysize=8.1cm \epsfbox{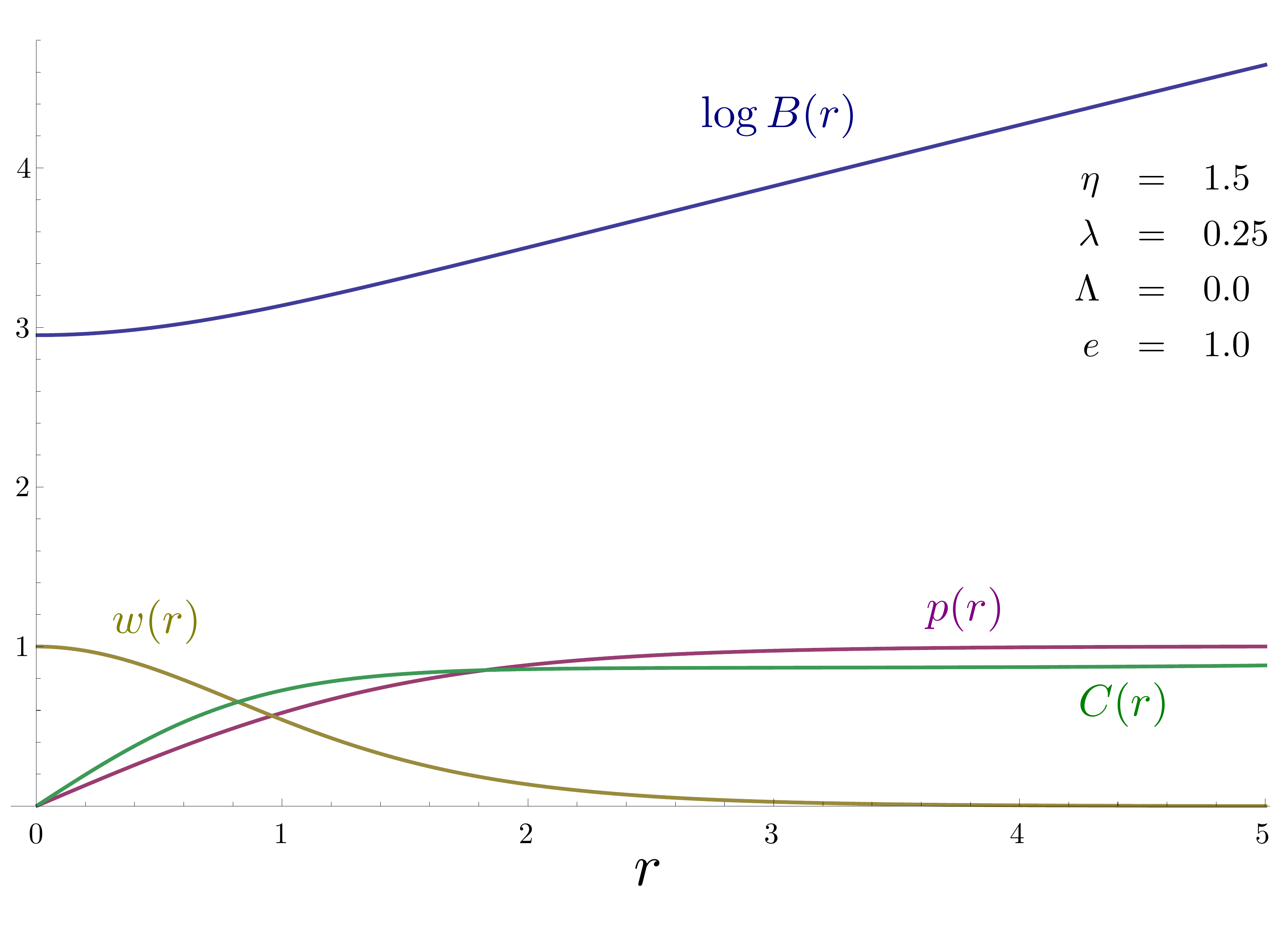}
\caption {A bubble of nothing in an AdS compactification. The core of
the monopole (surface of the bubble) is at $r = 0$, where the
$S^2$ degenerates.  The ``warp factor" $B(r)$ is nonzero at the core,
and grows exponentially toward the AdS boundary, where all fields
approach their vacuum values.  Throughout, we use reduced Planck units ($\kappa = 1$).}
\label{AdS4-BON}
\end{figure}
The first compactification we consider is to $AdS_4 \times S^2$, which 
occurs for all values of $n$ in a landscape with $\Lambda \le 0$, as well as 
for $n < e^2/(2\kappa^4 \Lambda)$, regardless of $\Lambda$. 
Bubbles of nothing were studied \cite{BPBS} in a much
simpler landscape whose vacua are all of this type. The minimal case is $\Lambda =0$, i.e., 
Freund-Rubin \cite{Freund}, which we begin with here.
In order to construct the bubble of nothing for this case, we
impose boundary conditions compatible with the asymptotic
compactification geometry.  Within our $SO(1,3)\times SO(3)$ invariant
metric ansatz Eq.~(\ref{eq:metric}), the asymptotically  $AdS_4 \times S^2$ solution is
\begin{figure}[htbp]
\centering\leavevmode
\epsfysize=7cm \epsfbox{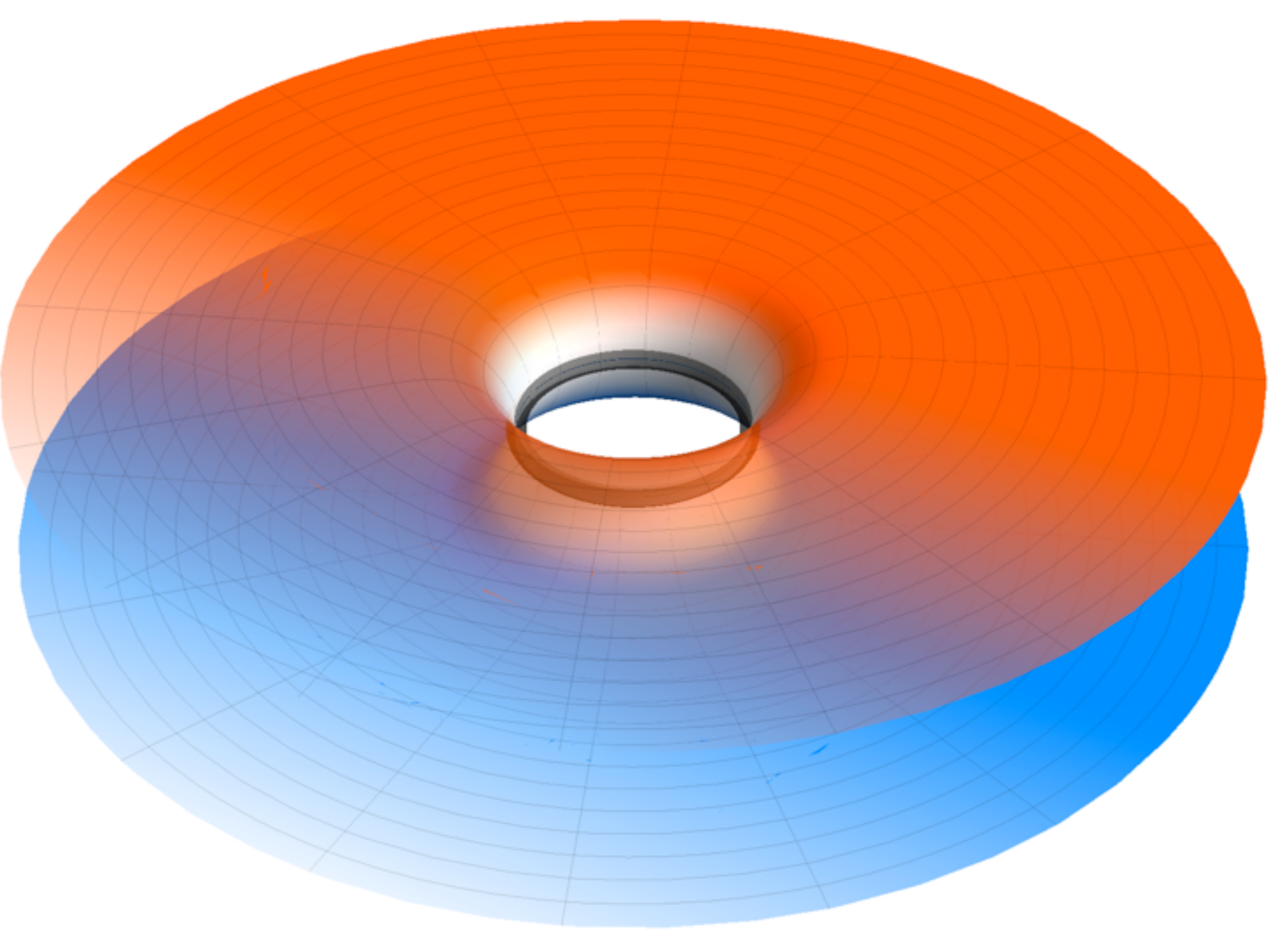}
\caption {Illustration of a bubble of nothing in an AdS
compactification. The vertical position represents the radion
$C(r)$, plotted as a function of the warp factor $B(r)$, represented
here by radial position.  The Euclidean $SO(4)$ or spatial $SO(3)$
symmetry is manifest in the rotational symmetry of the
illustration. Hue represents the wound electromagnetic flux (see
Fig.~\ref{n1compact}), and saturation is proportional to the scalar
field magnitude, $p(r)$. The thick black ring represents the position
of the defect.}
\label{EucAdSBON}
\end{figure}
\begin{eqnarray}
\label{adsasymptotic}
p(r)  &\to&  1\,, \hspace{3cm} w(r) \to 0\,, \nonumber\\
C(r)  &\to&  C_{\infty}\,, \hspace{1.4cm} B'(r)/B(r) \to \left|H\right|\,,
\end{eqnarray}
as $r\to\infty$.
The values of $\left|H\right|$ and $C_{\infty}$ for the $n=1$, $\Lambda=0$ case can
be seen in Eq.~(\ref{EYMH-landscape-solutions}) to be
\begin{eqnarray}
C_{\infty}  =  \sqrt{\frac{3 \kappa^2}{4 e^2}}\,,\hspace{2cm}
\left|H\right|  =  \sqrt{\frac{4 e^2}{27 \kappa^2}}~.
\end{eqnarray}

\begin{figure}[htbp]
\centering\leavevmode
\epsfysize=7cm \epsfbox{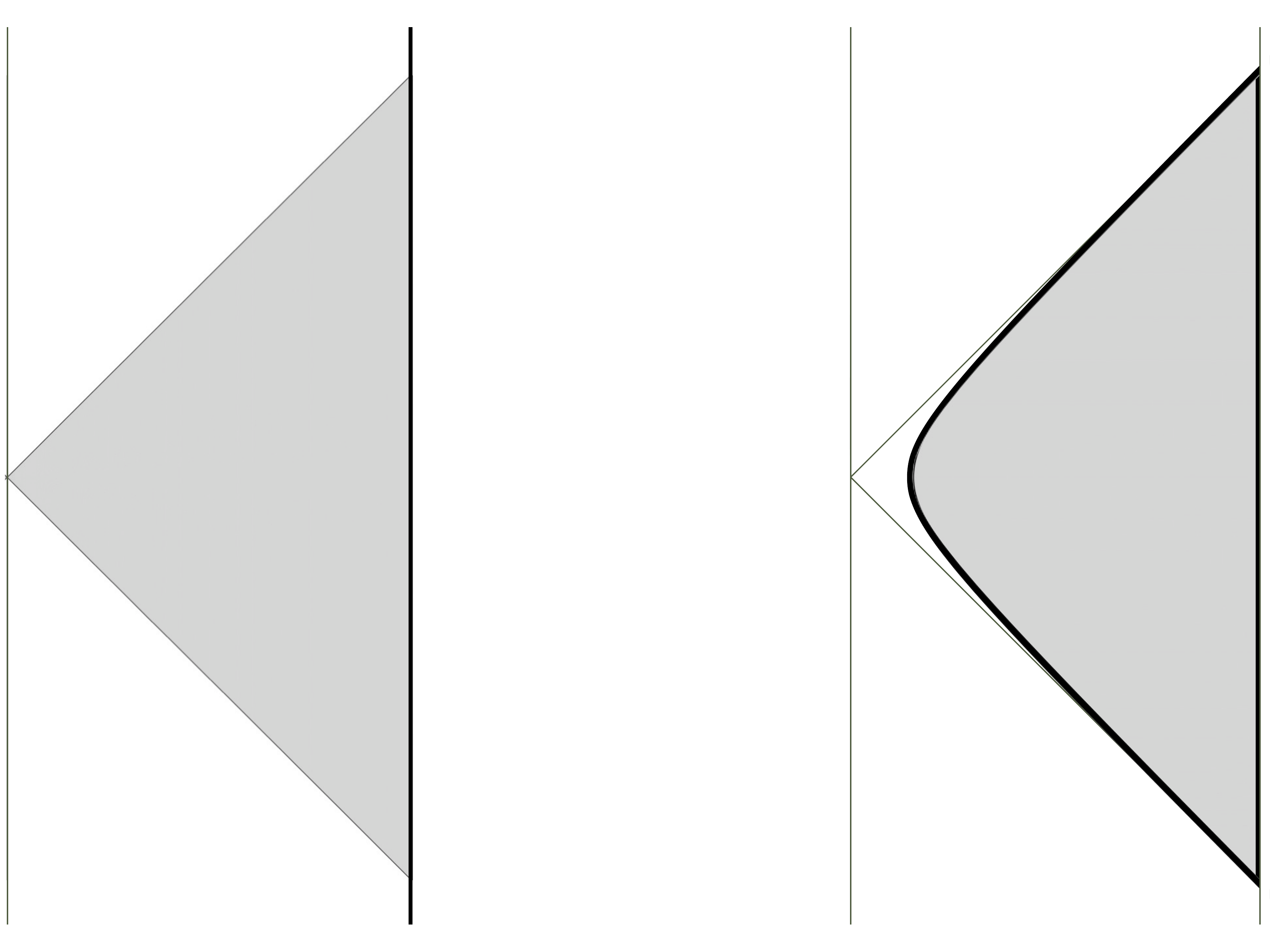}
\caption {Conformal diagrams of the AdS geometries.  Left:  The de Sitter slicing of $AdS_4$ covers the shaded
region.  Right: The bubble of nothing geometry only exists in the
shaded region outside the bubble wall denoted by the thick black line.} 
\label{confdiagAdS}
\end{figure}

The full solution, shown in Fig.~(\ref{AdS4-BON}), interpolates
between the near-core expansion given by Eq.~(\ref{nearcore}) and the
asymptotic solution Eq.~(\ref{adsasymptotic}).
The AdS bubble of nothing geometry is illustrated in Fig.~(\ref{EucAdSBON}),
which may be seen as the Euclidean solution, or as the spatial solution at the
moment of nucleation. After nucleation the bubble expands exponentially
eventually reaching the conformal boundary of the $4d$ anti-deSitter
space. This can be seen in Fig.~(\ref{confdiagAdS}) where we show the $4d$
conformal diagram of the bubble of nothing geometry for this $AdS_4$
compactification.\footnote{Note that all the conformal diagrams in 
this paper describe the $4d$ part of the geometry in Eq.~(\ref{eq:metric}).  Every interior point represents a large $S^2$ from the $4d$ part of
the spacetime times the small $S^2$ compactification manifold.} 

As a six dimensional geometry, one can interpret the behavior of the
warp factor $B(r)$ as indicating the presence of a throat-like region in our 
spacetime. The gravitational potential due to the 
warp factor reveals the (in this case) attractive nature of the bubble
geometry ($B(r)$ is decreasing toward the bubble). Gravitationally attractive 
throats appear in the context of warped compactifications \cite{GKP}.
A gapped warped throat (e.g., Klebanov-Strassler) in global
coordinates may even be thought of as a bubble of nothing
geometry, albeit with cylindrical rather than de Sitter isometry, and
lacking a negative mode.

In our solutions, the bubble wall represents the smooth termination of this 
throat. We can use the gravitational properties of the throat as a proxy for
the effective $4d$ tension of the bubble. Since a bubble of nothing
accelerates toward an outside observer, the throat is attractive, and
the apparent $4d$ tension of the domain wall is negative \cite{Garriga:1998ri,Garriga:1998tm}.  
One can see this by calculating the effective tension one
would have to place in a $4d$ spacetime to orbifold two identical 
copies of the shaded region in Fig.~(\ref{confdiagAdS}).

\subsection{The decay of ${\mathbb R}^{1,3} \times S^2$ spacetimes}

\begin{figure}[htbp]
\centering\leavevmode
\epsfysize=8cm \epsfbox{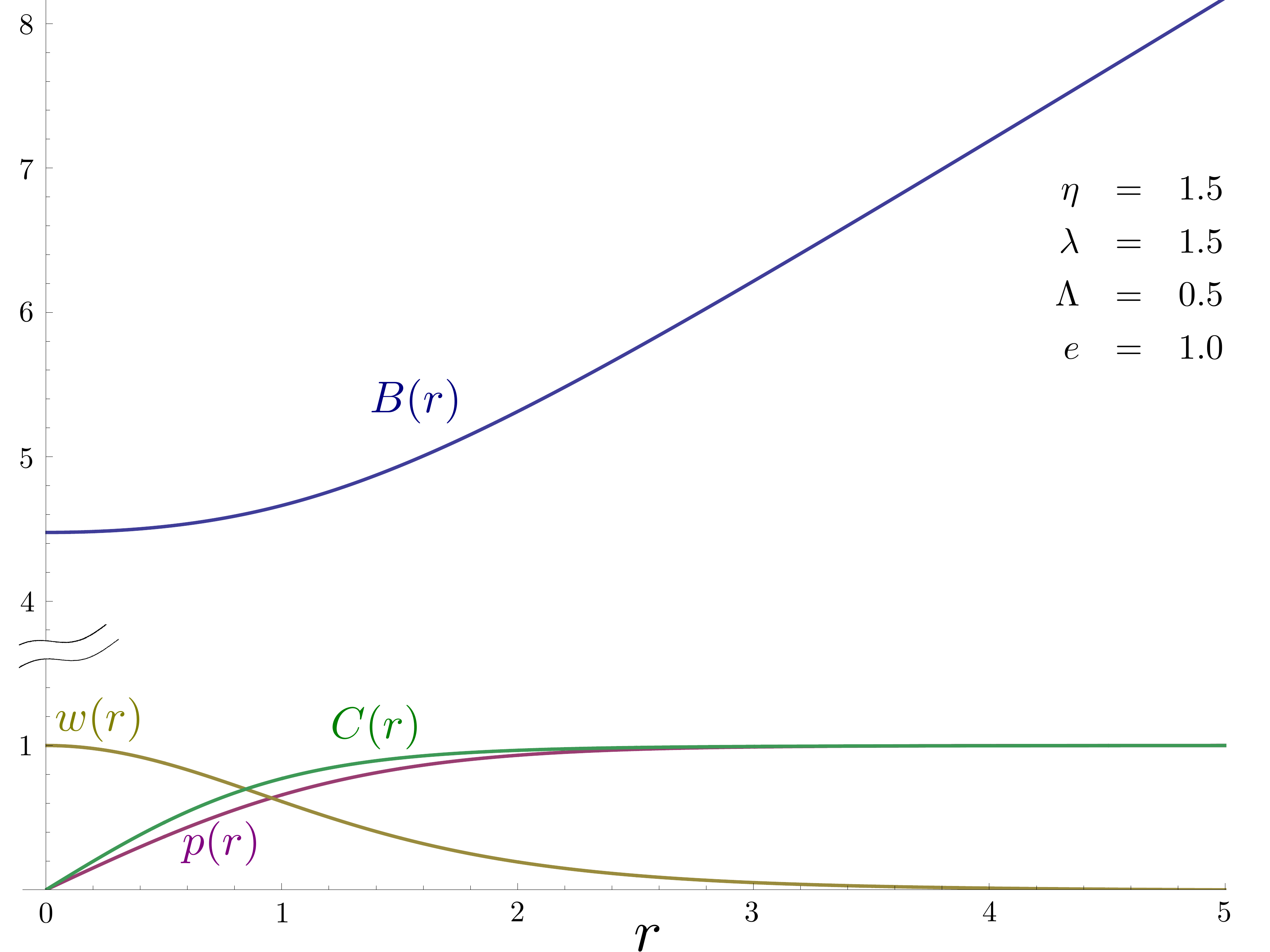}
\caption {A bubble of nothing in a Minkowski compactification}
\label{M4-BON}
\end{figure}

\begin{figure}[htbp]
\centering\leavevmode
\epsfysize=7cm \epsfbox{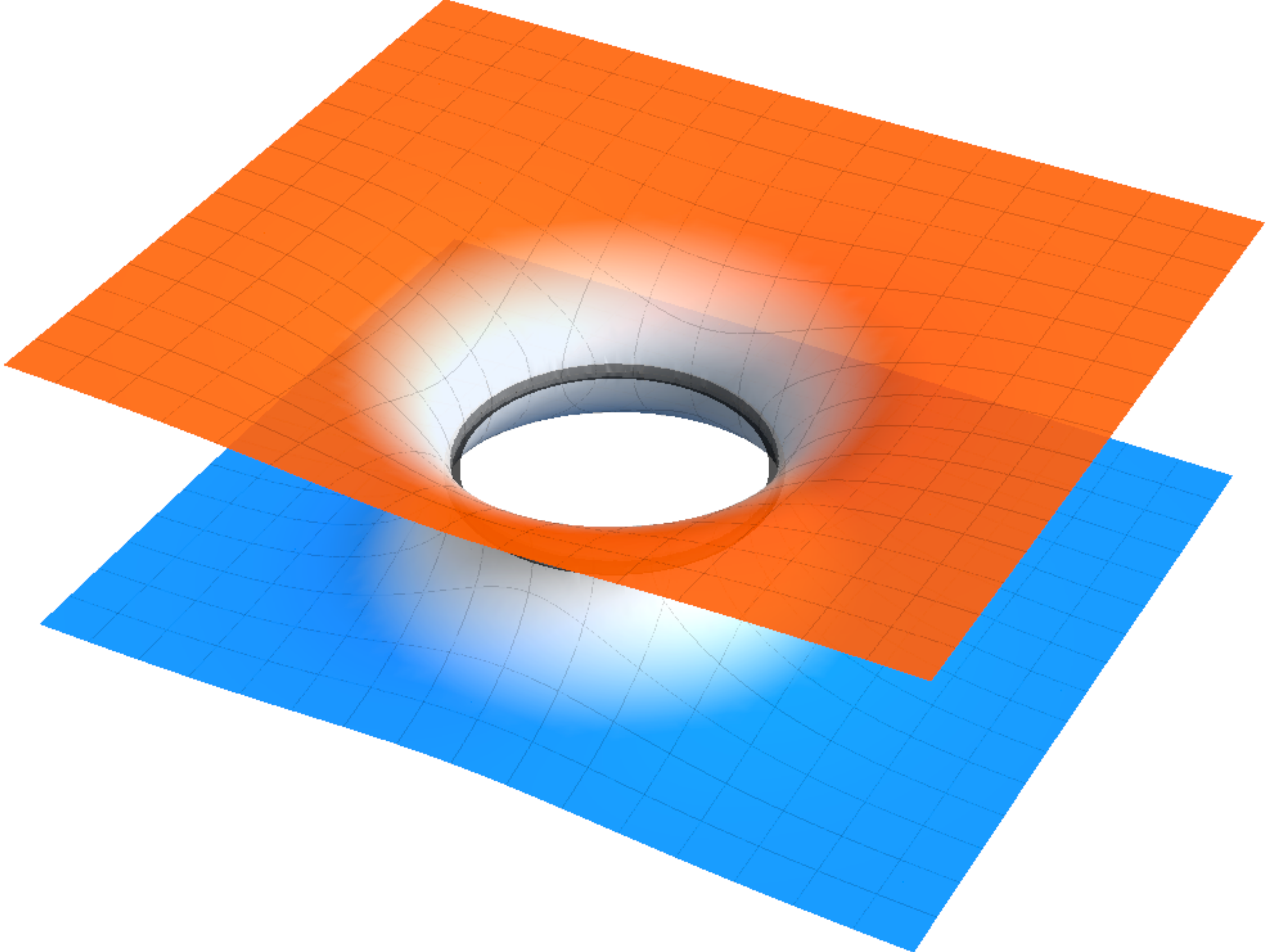}
\caption {A bubble of nothing in a Minkowski compactification. The
vertical position represents the radion $C(r)$, plotted as a
function of the warp factor $B(r)$,  whose minimum occurs at the
core of the defect (thick dark ring)}
\label{eucflatbon}
\end{figure}

We may uplift the effective $4d$ cosmological constant to zero
for the $n=1$ vacua by raising the $6d$ cosmological
constant to
\begin{equation}
\Lambda = \frac{e^2}{2 \kappa^4}  ~~.\nonumber\\
\end{equation}
The asymptotic solution is then given by
\begin{figure}[htbp]
\centering\leavevmode
\epsfysize=8cm \epsfbox{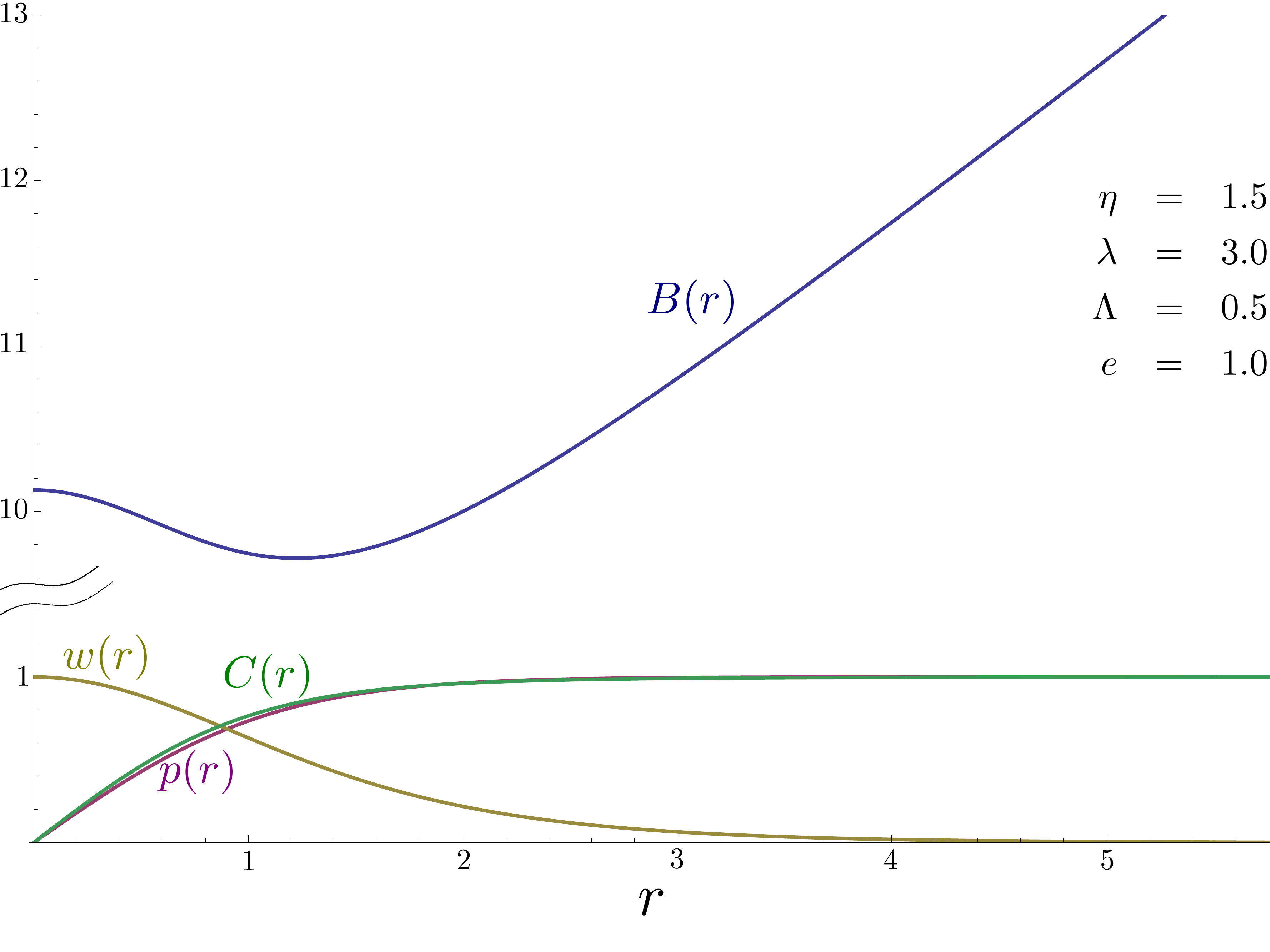}
\caption {A punted bubble of nothing in a Minkowski
  compactification. The minimum of the warp factor $B(r)$ does not occur at the core.}
\label{punt-M4-BON}
\end{figure}

\begin{eqnarray}
p(r)  &\to&  1\,, \hspace{3cm} w(r) \to 0\,, \nonumber\\
C(r)  &\to&  C_{\infty}= \frac{\kappa}{e}\,,  \hspace{1.6cm} B'(r) \to 1\,,
\end{eqnarray}
as $r\to\infty $ .
A numerical example of the bubble of nothing geometry in this vacuum is shown in
Fig.~(\ref{M4-BON}).

\begin{figure}[htbp]
\centering\leavevmode
\epsfysize=8cm \epsfbox{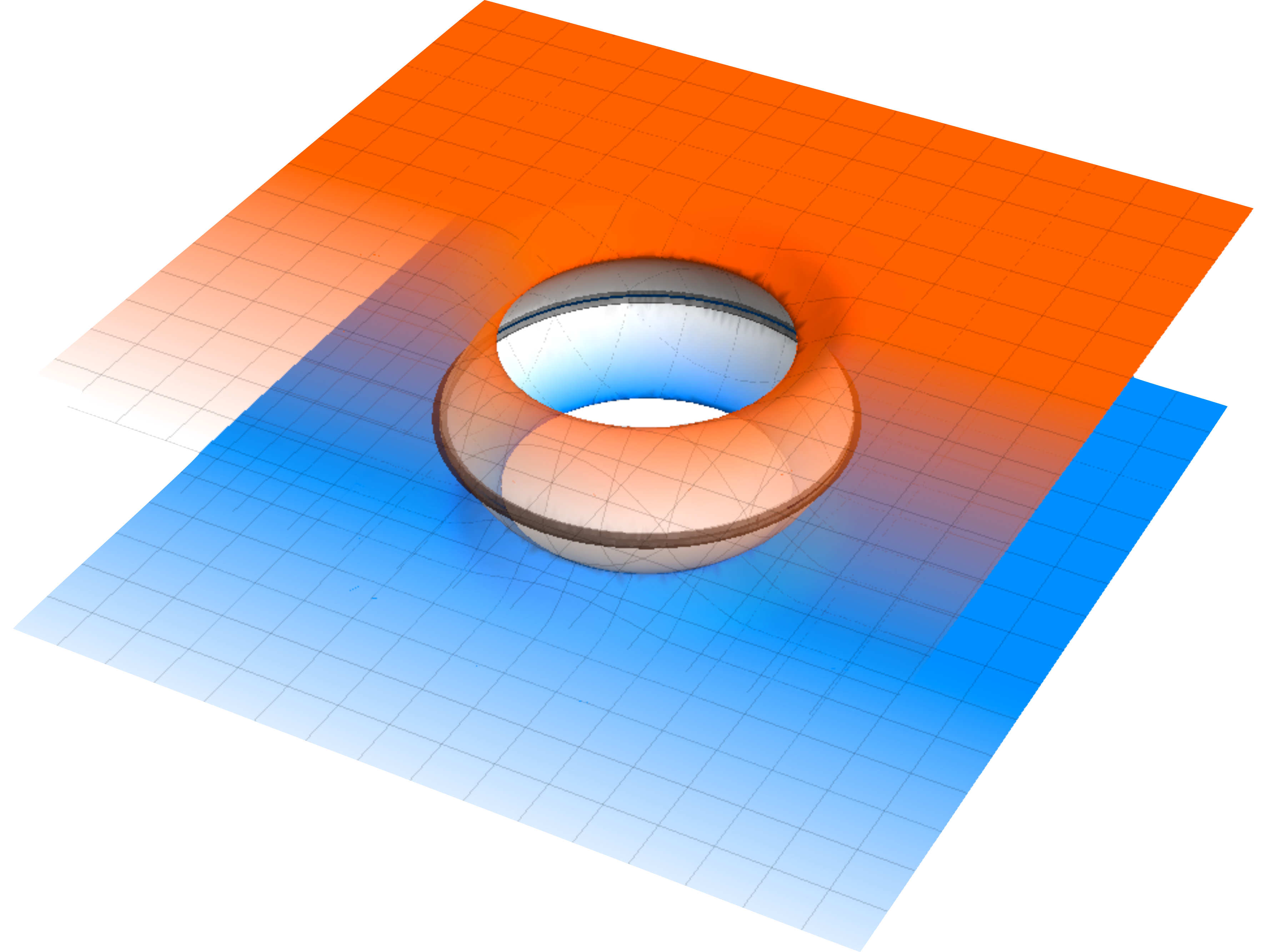}
\caption {Illustration of a punted bubble of nothing in Minkowski
  space.  The vertical separation represents the radion $C(r)$, plotted
  as a function of the warp factor $B(r)$,  whose minimum occurs away
 from the core of the defect (thick dark ring).}
\label{eucflatpuntbon}
\end{figure}

As mentioned before, we have introduced new parameters $\lambda$ and
$\eta$ into our model which affect local properties of the bubble
wall, but which are independent from the asymptotic solution. In
Figs.~(\ref{punt-M4-BON}\,-\,\ref{eucflatpuntbon}) we give an example of such variation by
finding a new type of bubble solution that is qualitatively different
in the near tip region.  

There, the warp factor $B(r)$ displays a punt
shape, like a wine bottle achieving its minimum slightly
away from the core. We can understand the existence of this family
of solutions demonstrating the different form of the warp factor 
as a competition between the two contributions to the $4d$ gravitational properties
in this region, one coming from the bubble of nothing itself, and the
other from the magnetic 2-brane located at the surface of the bubble.

\subsection{The decay of $dS_4 \times S^2$ spacetimes}

\begin{figure}[htbp]
\centering\leavevmode
\epsfysize=7cm \epsfbox{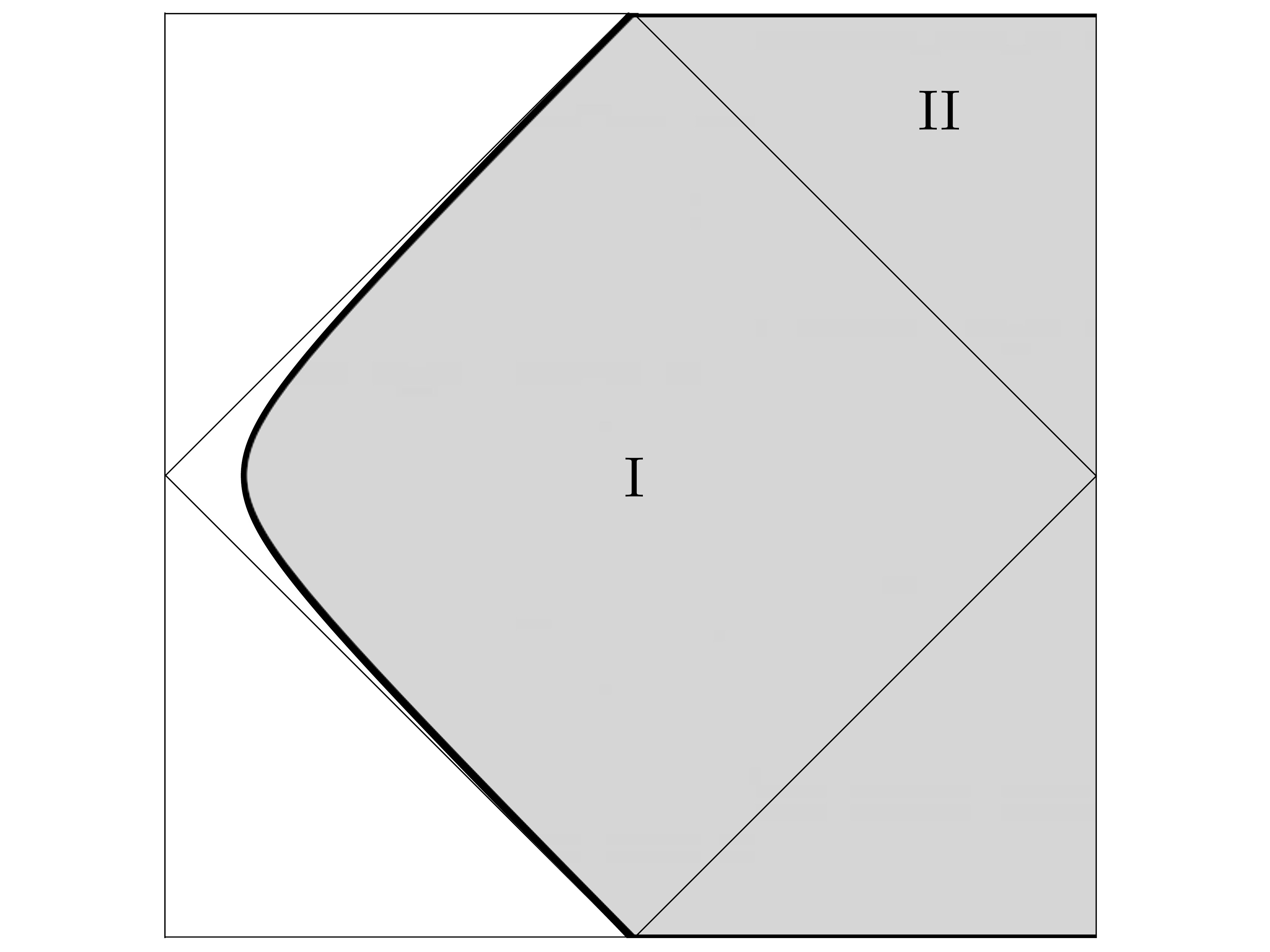}
\caption {A conformal diagram for a bubble of nothing in $dS_4$.  The
  spacetime only exists in the shaded region.}
\label{confdiagdSBON}
\end{figure}

Bubbles of nothing in de Sitter space are complicated by the existence
of a cosmological horizon\footnote{For a critique of exponential decay
of de Sitter vacua, see \cite{Shlaer:2009vg}.}. The bubble
geometry in this case has an exterior region of finite radius, $0 < r < r_h$, 
where $r_h$ is defined by $B(r_h)=0$. This is denoted by the horizon which bounds region I in Fig.~(\ref{confdiagdSBON}). Expanded about the
horizon at $r = r_h$, the solution takes the form
\begin{eqnarray}
\label{dshorizon}
p(r) & = & p_h + p_2 (r-r_{h})^{2} + \cdots\,, \nonumber\\
w(r) & = & w_h + w_2 (r-r_{h})^{2}\cdots \,,\nonumber\\
B(r) & = & (r-r_{h}) - B_3(r-r_{h})^{3} + \cdots \,,\\
C(r) & = & C_{h} - C_2 (r-r_{h})^{2} +\cdots \,,\nonumber
\end{eqnarray}
where $p_2, w_2, B_3$, $C_2$, etc. can be found in terms of the three
field values at the horizon, $p_{h}, w_{h}$, and $C_{h}$. We relegate
the more complete expressions for these expansions to Appendix \ref{appendix-expansion-horizon}.

Unlike the asymptotically flat or AdS bubbles of nothing, there is
no topological distinction between a bubble of nothing in $dS_4\times
S^2$ and other physical solutions, including $dS_6$. Intuitively, a
bubble of nothing should be a boost-invariant solution with a
degenerating extra-dimensional fiber, which within our ansatz is a zero for the function $C(r)$.
The broadness of these criteria becomes apparent when one considers the anisotropic slicing
of $dS_6$, whose metric is given by $B(r) = \cos r$, $C(r) =  \sin r$:
\beq
ds^2 = \cos^2(r) (-dt^2 + \cosh^2t~d\Omega_2^2) + dr^2 + \sin^2r~d\Omega_2^2\,.\label{aniso-dS6}
\eeq
Remarkably, this appears to be a bubble of nothing\footnote{A $5d$ version of this interpretation appears in \cite{Garriga:1998ri}.}.
In this case, any observer is on the core of the bubble at $r = 0$ and sees a
cosmological horizon at $r = \pi/2$, with topology given by an $S^2\times S^2$ fibration
of $S^4$.

As this example demonstrates, we must adopt a more restrictive definition
if we demand the bubble of nothing describe a decay
channel for flux compactifications. We will
therefore look not only for boost-invariant solutions with a smooth core region 
but also solutions with an asymptotic region which approaches a $4d$ 
flux vacuum. This requires us to determine the behavior 
of the solutions beyond the cosmological horizon, in what we denote by 
region II of Fig.~(\ref{confdiagdSBON}). One can do this by
analytically continuing the metric ansatz across this horizon 
via the substitution $r\rightarrow it$ and $t\rightarrow
\chi+i\pi/2$ in Eq.~(\ref{eq:metric}), yielding
\begin{equation}
ds^{2}= - dt^{2} + B^{2}(t)d{\cal{H}}_{3}^{2} + C^{2}(t)d\Omega_{2}^{2}~,
\label{timelikeeq}
\end{equation}
where $d{\cal{H}}_{3}^{2}$ is the unit metric on three dimensional hyperbolic space, 
\begin{equation}
d{\cal{H}}_{3}^{2}=d\chi^{2}+\sinh^{2}\chi d\Omega_{2}^{2}~.
\end{equation}
Matter fields in this region are given by
\begin{eqnarray}
\Phi^{a}(t)&=&\eta\ p(t) (\sin\theta \cos\varphi, \sin\theta
\sin\varphi, \cos\theta)\,,\nonumber\\
A^{a}_{\mu}(t)&=&A^{a}_{r}(t)=0\,,\nonumber\\
A^{a}_{\theta}(t)&=&\frac{1-w(t)}{e} (\sin\varphi,-\cos\varphi,0)\,,\\
A^{a}_{\varphi}(t)&=&\frac{1-w(t)}{e} \sin\theta (\cos\theta
\cos\varphi, \cos\theta \sin\varphi, -\sin\theta)~.\nonumber
\end{eqnarray}
The general expansion of the fields about the light-cone
($t = 0$) yields
\begin{eqnarray}
p(t) & = & p_h - p_2 t^{2} + \cdots\,,\nonumber\\
w(t) & = & w_h - w_2 t^{2}\cdots\,,\nonumber\\
B(t) & = & t + B_3 t^{3} + \cdots\,,\\
C(t) & = & C_h + C_2 t^{2} +\cdots \nonumber~,
\end{eqnarray}
where the three undetermined coefficients $p_h, w_h,$ and $C_h$ are
trivially related to the field values across the horizon ($r = r_h$).

\begin{figure}[htbp]
\centering\leavevmode
\epsfysize=8cm \epsfbox{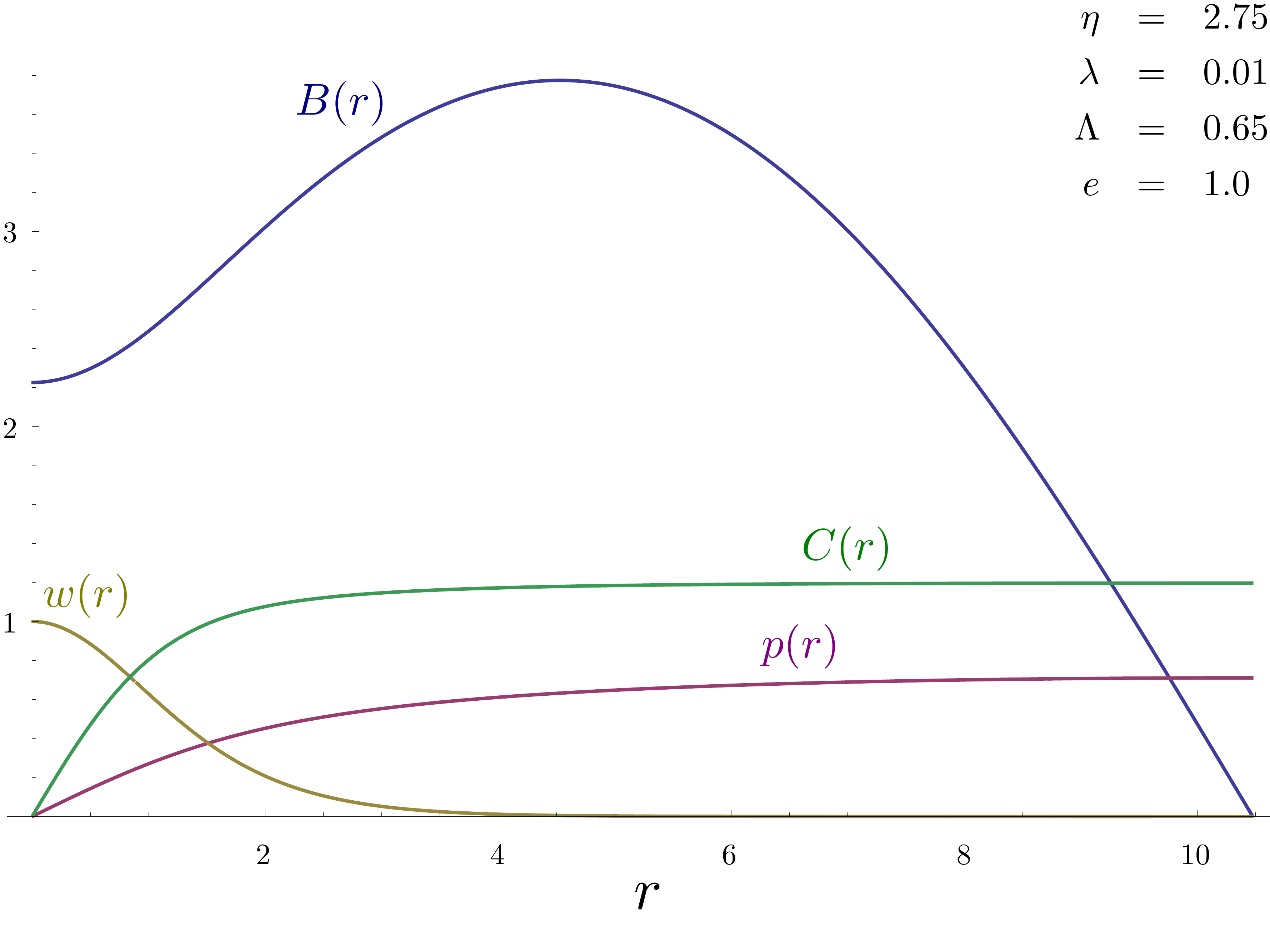}
\caption {A bubble of nothing in a de Sitter compactification.  The
  solution resembles a conventional bubble of nothing in the near-core region.}
\label{ds-bon-graph}
\end{figure}

\begin{figure}[htbp]
\centering\leavevmode
\epsfysize=8cm 
\epsfbox{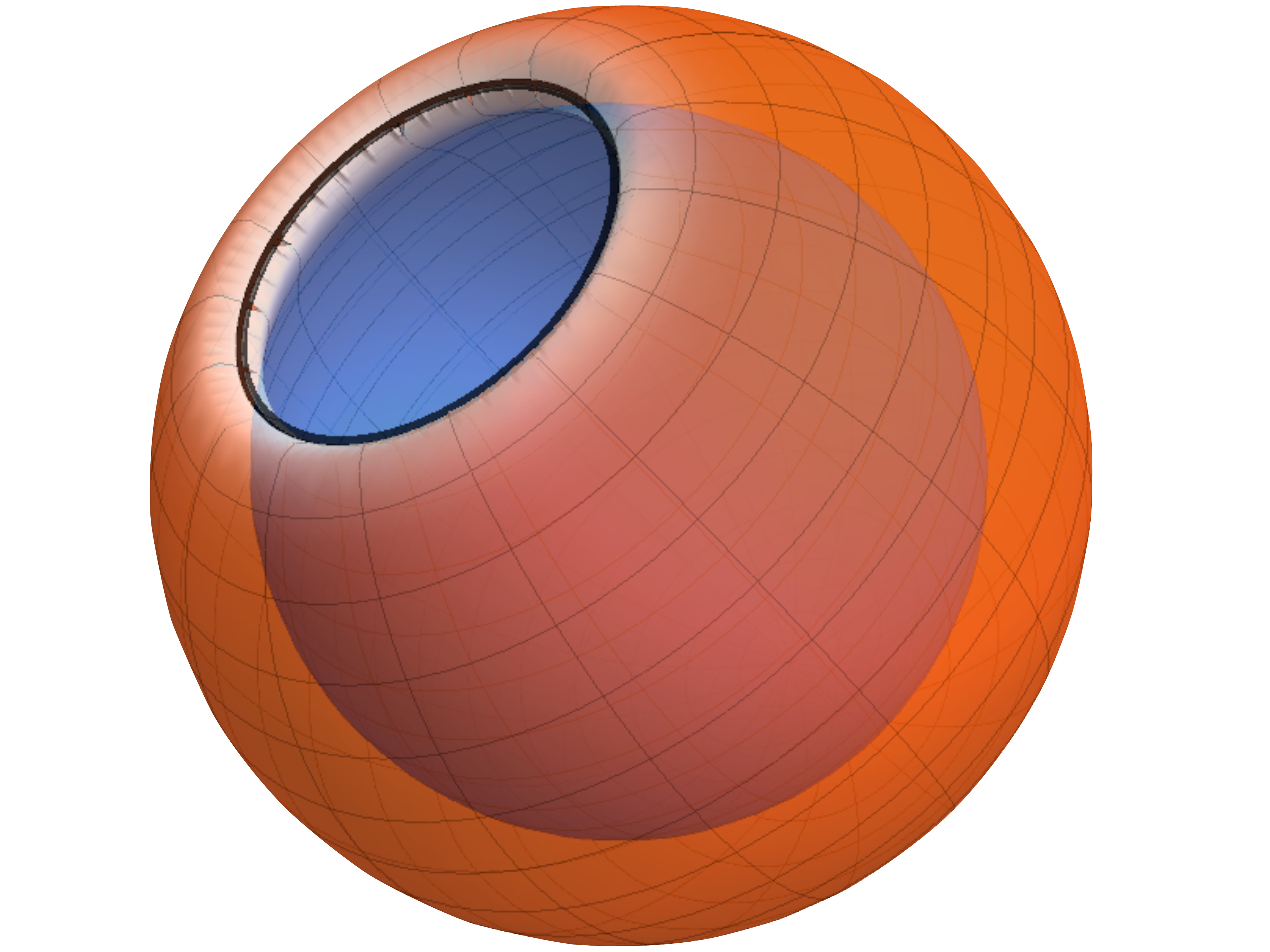}
\caption {A bubble of nothing in $dS_4\times S^2$. This illustrates
either the Euclidean solution, with cosmological horizon antipodal
to the hole, or the global spatial solution at the moment of nucleation.}
\label{ds-bon}
\end{figure}

\begin{figure}[htbp]
\centering\leavevmode
\epsfysize=12cm \epsfbox{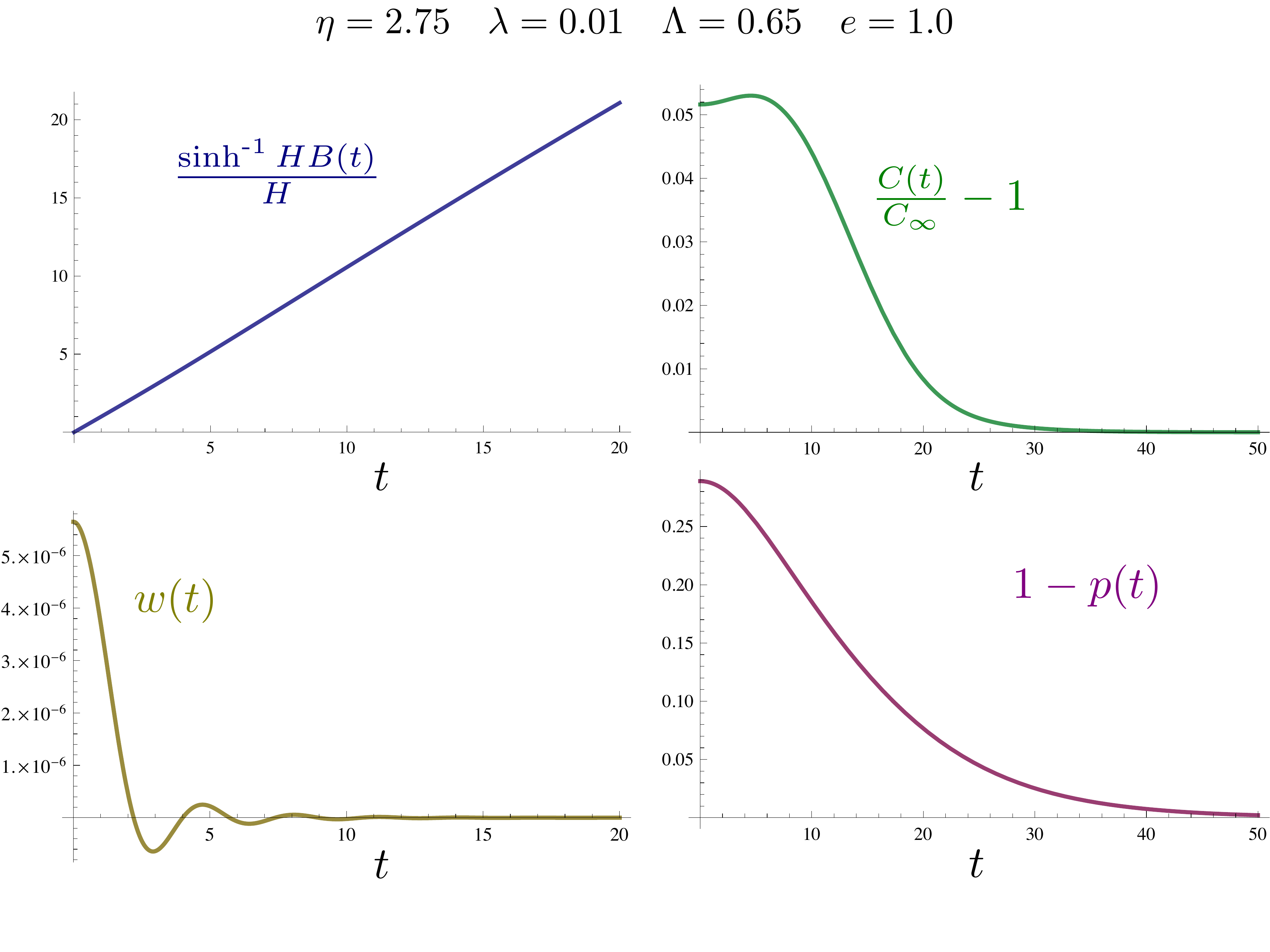}
\caption {The future-directed evolution of the fields in region I of Fig.~(\ref{confdiagdSBON}).  The solution asymptotes to the $4d$ flux vacuum as $t\to\infty.$}
\label{time-ds-bon-graph}
\end{figure}

Using the time-continued equations of motion shown in Appendix~\ref{appendix-timelike-eqs},
we numerically integrate the solution forward in time,
taking as initial conditions the values of the fields at the horizon
separating the future region (II) from the spacelike region (I). 

For certain parameter values, one can find
solutions for the bubble of nothing asymptotic to $dS_4 \times S^2$.
Shown in Fig.~(\ref{ds-bon-graph}\,-\,\ref{ds-bon}) is a numerical solution in
region I, where the behavior $B(r)\rightarrow 0$ signals the
appearance of a cosmological horizon, and the remaining functions 
behave as described in Eq.~(\ref{dshorizon}). Following the procedure
outlined above, one can find that indeed this solution relaxes to 
the appropriate $dS_4 \times S^2$ compactification. We show in
Fig.~(\ref{time-ds-bon-graph}) the posterior evolution
of the fields beyond the horizon in region II.

There is however, a different class of solution one can find in this
future directed region.  There are solutions which lead to runaway behavior for the radion
$C(t)$. This is manifestly different from a compactification. In fact, 
it is not difficult to see that at late times this geometry asymptotes to 
six dimensional de Sitter space written in an anisotropic gauge. The 
interpretation of these types of solutions is not as a bubble of
nothing in a flux compactification, but as an instanton describing 
the creation of smooth magnetically charged 2-branes in $dS_6$ \cite{BPSPV-2}.
We give an example of such type of solutions in Appendix
\ref{appendix-brane-dS6}.

An interesting feature of the bubble of nothing in de Sitter
compactifications is the symmetry between the `excised' region and the
undisturbed region, as can be seen in Figs.~(\ref{confdiagdSBON}) and
(\ref{ds-bon}). In fact, within the family of solutions are those
where the excised region is far larger than the undisturbed region. In
the framework of this paper, this should be interpreted as the
spontaneous collapse of a super-horizon sized region to nothing.  Another interpretation of this solution is that of an
instanton describing the spontaneous {\em creation} of an open flux
compactification. The ambiguity between these interpretations disappears when
considering the analogous solutions in flat and AdS
compactifications. We therefore refer to these solutions 
as {\em bubbles from nothing} \cite{BPRStoappear} .

\section{Conclusions}
We have demonstrated a new instability of flux compactifications, and
a new topology for the bubble of nothing. Like the original bubble of nothing
of the Kaluza--Klein vacuum, the bubbles we present are smooth
gravitational instantons asymptotic to a compactification
geometry. The principal 
new ingredients are
\begin{itemize}
\item The bubble surface is charged with
respect to the flux employed to stabilize the compactification.
\item The solutions describe the smooth degeneration
of an $S^2$, rather than the previously known $S^1$ cases.
\item The bubbles have a variety of $4d$ effective tensions, which can
  be negative or positive.
\item The solutions preserve the isometry of the compactification
  manifold only for flux number $n = \pm1$.
\item The instability may occur for perturbatively stable flat, AdS,
  or dS compactifications, although
it does not exist for all parameter values $\eta$, $\lambda$.
\end{itemize}
A consequence of the topology of the bubbles of nothing presented
here is that spin structure cannot play a role
in excluding the instability; every configuration considered here
satisfies $\pi_1({\cal{M}}_6) = 0$.

Families of solutions exist for anti de Sitter, Minkowski, as well as
de Sitter compactifications, although the taxonomy of a bubble of
nothing in $dS_4\times S^2$ is complicated by the lack of a
topological distinction from other solutions which
are physically distinct (e.g., defects in $dS_6$ and the {\em bubble from nothing}).
Roughly speaking, a bubble of nothing should be gravitationally
attractive over a large range of distances, meaning an observer must accelerate away from the bubble
in order to avoid collision. In this case, the effective $4d$ tension
of the boundary of spacetime is negative. Because of the natural de
Sitter slicing, the throat picture of a bubble of nothing is
reminiscent of the dS/dS correspondence \cite{Alishahiha:2004md}.

On the other hand, spin structure does not allow one to project out
the bubble of nothing instability in this more general case. It would therefore
be interesting to find what mechanism forbids the decay of  
supersymmetric flux vacua to nothing.

By increasing the monopole parameters $\eta$ and $\lambda$, one can
preserve the long range attractive nature of the solution despite a
short range gravitational repulsion, as shown in
Fig.~(\ref{eucflatpuntbon}). 

A more drastic solution, to be discussed
in a future publication \cite{BPRStoappear}, describes a purely
repulsive boundary, which we have referred to as a ``{\em bubble from
nothing}."  Although this solution is topologically equivalent to a
bubble of nothing in de Sitter compactifications, the corresponding
interpretation (certainly in the flat and AdS case) is
distinct from that of a bubble of nothing. 

\acknowledgements
J.~J.~B-P and B.~S.~would like to thank the Yukawa Institute for hospitality and support during
the Gravity and Cosmology GC2010 workshop, where part of this work was
completed. We would also like to thank Roberto Emparan, Jaume Garriga,
Ken Olum, Oriol Pujolas, Mike Salem and Alex Vilenkin for helpful discussions.
J.~J.~B.~-P.~  is supported 
in part by the National Science Foundation under grant 0653361.  B.~S.
is supported in part by Foundational Questions Institute grant RFP2-08-26A.

\appendix

\section{Compactification Solutions for $n = 1$}
\label{appendix-n-equal-1}
For compactifications of unit flux number, Eqs.~(\ref{eq:einstein}),
(\ref{eq:higgs}), and (\ref{eq:yangmills}) reduce to
\begin{eqnarray}
3H^{2}+\frac{1}{C^{2}} &=& \kappa^{2}\bigg(\frac{\eta^2 p^{2}w^{2}}{C^{2}} +
\frac{(1-w^{2})^{2}}{2e^{2}C^{4}} + \frac{\lambda
  \eta^4}{4}(p^{2}-1)^{2} + \Lambda\bigg)\,,\nonumber\\
6H^{2} &=& \kappa^{2}\bigg(-\frac{(1-w)^{2}}{2e^{2}C^{2}} +
\frac{\lambda \eta^4}{4}(p^{2} - 1)^{2}+\Lambda\bigg)\,,\nonumber\\
0 &=&\frac{2 p w^{2}}{C^{2}}+\lambda \eta^2\ p\ (p^{2}- 1)\,,\nonumber\\
0 &=&\frac{w(1-w^{2})}{C^{2}}-e^{2}\eta^2 p^{2}w\,.
\end{eqnarray}

The solution to these equations is not unique. They can be categorized
by their stability, as below.
\subsection{Stable Compactification Solutions}
 
Stable solutions exist when the fields relax to their vacuum values,
$p=1$, $w=0$, yielding
\begin{eqnarray}
H^{2} & = & \frac{2\kappa^{2}\Lambda}{9} -
\frac{2e^{2}}{27\kappa^{2}}\bigg(1 + \sqrt{1 -
  \frac{3\kappa^{4}\Lambda}{2e^{2}}}\bigg)\,,\nonumber\\
C^{2} & = & \frac{1}{\kappa^{2}\Lambda}\bigg(1 - \sqrt{1 -
  \frac{3\kappa^{4}\Lambda}{2e^{2}}}\bigg)\,.
\end{eqnarray}
One can see that this is a stable configuration by looking at the $4d$
effective action about this solution. (See the
main part of the text for a discussion on this point.) 
These are the most interesting solutions for our purpose, although
one can find several other solutions which are unstable. 

\subsection{Unstable Compactification Solutions}

For completeness, we construct unstable configurations which exist in the
$n=1$ case.\footnote{The first type of unstable solutions presented here
may also occur for $n>1$, although none of the subsequent examples
generalize in this way.}  Our numerical solutions approach only stable
configurations, as should be the case for purely non-perturbative instabilities.

\subsubsection{Compactifications with $p=1$ and $w=0$}

Here the matter fields have relaxed to their respective vacua, but $C$
is sitting at an unstable equilibrium for the size of the
compactification manifold. In other words, these solutions are the straightforward 
generalization of the Nariai compactification solutions. The
solutions take the form

\begin{eqnarray}
H^{2} & = & \frac{2\kappa^{2}\Lambda}{9} -
\frac{2e^{2}}{27\kappa^{2}}\bigg(1 - \sqrt{1 -
  \frac{3\kappa^{4}\Lambda}{2e^{2}}}\bigg)\,,\nonumber\\
C^{2} & = & \frac{1}{\kappa^{2}\Lambda}\bigg(1 + \sqrt{1 -
  \frac{3\kappa^{4}\Lambda}{2e^{2}}}\bigg)\,.
\end{eqnarray}

\subsubsection{Compactification with $p=0$ and $w=1$}

These configurations are clearly unstable since vanishing $p$ implies that
the scalar triplet sits at the top of its potential.  Here,

\begin{eqnarray}
H^{2}&=&\frac{\kappa^{2}}{24}(\lambda\eta^{4}+4\Lambda),\nonumber\\
C^{2}&=&\frac{8}{\kappa^{2}(\lambda\eta^{4}+4\Lambda)}.
\end{eqnarray}

\subsubsection{Non-zero constant $p$-$w$ Solutions}
  
This is a solution where both $p$ and $w$ are non-zero constants
different from their vacuum values. We have checked numerically that this 
type of solution is unstable to decompactification, as was indicated in \cite{Clement}.

To simplify the notation, we define the quantity
\begin{equation}
\alpha = \sqrt{(8e^{2} + 4\lambda(\eta^{2}\kappa^{2} - 1))^{2} -
  24\kappa^{2}(e^{2}(\lambda\eta^{4} + 4\Lambda)-2\lambda\Lambda)},
\end{equation}
allowing the solutions to be written
\begin{eqnarray}
p &=& \sqrt{\frac{-4e^{2} +
    2\lambda+\lambda\eta^{2}\kappa^{2}\pm\frac{1}{2}\alpha}{(3\lambda
    - 6e^{2})}}\,,\nonumber\\
w &=& \sqrt{\frac{\lambda\bigg(43^{2}\eta^{2} +
    2\kappa^{2}\lambda\Lambda+e^{2}(\lambda\eta^{4}\kappa^{2} -
    2\lambda\eta^{2} -
    4\kappa^{2}\Lambda)\bigg)\pm\frac{1}{2}e^{2}\eta^{2}\alpha}{\kappa^{2}(2e^{2}
    - \lambda)\bigg(e^{2}(\lambda\eta^{4}+4\Lambda)-2\lambda\Lambda\bigg)}}\,,\nonumber\\
H^{2} &=& \frac{\kappa^{2}\bigg(e^{2}(\lambda\eta^{4} + 4\Lambda) -
  2\lambda\Lambda\bigg)\bigg(2e^{2} -
  \lambda+\eta^{2}\kappa^{2}\lambda\pm\frac{1}{2}\alpha\bigg)}{9(2e^{2}
  - \lambda)\bigg(4e^{2} - 2\lambda +
  2\eta^{2}\kappa^{2}\lambda\pm\frac{1}{2}\alpha\bigg)}\,,\\
C^{2} &=& \frac{6\kappa^{2}}{4e^{2} - 2\lambda +
  2\lambda\eta^{2}\kappa^{2}\mp\frac{1}{2}\alpha}\,.\nonumber
\end{eqnarray}

\section{Expansion about the soliton core}
\label{appendix-expansion-core}
The expansion of the equations of motion takes the following
form for the most general smooth solution describing the core of the magnetically
charged inflating 2-brane.  In the ansatz of Eqs.~(\ref{eq:metric}-\ref{eq:matter}),
\begin{eqnarray}
p(r) & = & p_{1} r + \cdots\nonumber\,,\\
w(r) & = & 1 + w_{2}r^{2} + \cdots\nonumber\\
B(r) & = & B_{0} + \frac{B_{0}}{12}\bigg[\frac{1}{B_{0}^{2}} 
+ \frac{3(2e^{2} + B_{0}^{2}e^{2}\eta^{2}p_{1}^{2}\kappa^{2} 
+ 8 B_{0}^{2}w_{2}^{2}\kappa^{2})}{2 B_{0}^{2}e^{2}}\\
&& - \frac{\kappa^{2}}{4}\left(6\eta^{2}p_{1}^{2} 
+ \eta^{4}\lambda + \frac{24w_{2}^{2}}{e^{2}} + 4\Lambda\right)\bigg]r^{2}
+ \cdots\,,\nonumber\\
C(r) & = & r - \frac{(2e^{2} + B_{0}^{2}e^{2}\eta^{2}p_{1}^{2}\kappa^{2}
+ 8 B_{0}^{2}w_{2}^{2}\kappa^{2})}{12 B_{0}^{2}e^{2}}r^{3} + \cdots\,.\nonumber
\end{eqnarray}
This expansion depends on three locally undetermined constants, $B_{0}$, $p_{1}$, and $w_{2}$, 
in terms of which all subsequent terms in the near-core expansions may be specified.

\section{Expansion about the cosmological horizon}
\label{appendix-expansion-horizon}

Some of the solutions we obtain possess a cosmological horizon,
defined by $B(r_h) = 0$. The general form for
a smooth expansions about the horizon $r=r_{h}$ is
\begin{eqnarray}
p(r) & = & p_h + p^h_2 (r-r_h)^{2} + \cdots,\nonumber\\
w(r) & = & w_h + w^h_2 (r-r_h)^{2}\cdots,\nonumber\\
B(r) & = & (r-r_h) - B_3^h(r-r_h)^{3} + \cdots,\\
C(r) & = & C_h - C^h_2 (r-r_h)^{2} +\cdots \nonumber ~,
\end{eqnarray}
where the coefficients of this expansion are given by
\begin{eqnarray}
p^h_2 & = & \frac{p_h\left[2 w_h^2 - \lambda \eta^2 C_h^2(1 - p_h)\right]} {8C_h^2},\nonumber\\
w^h_2 & = & \frac{w_h\left[w_h(w_h + e^2 \eta^2 p_h^2 C_h^2) -  1\right]} {8C_h^2},\nonumber\\
B^h_3 & = & \frac{1}{288C_h^4 e^2}\bigg\{10(w_h^2 - 1)^2 \kappa^2 + 8e^2 C_h^2(\eta^2 p_h^2 w_h^2
\kappa^{2}-1) - e^2 C_h^4 \kappa^2\left[\eta^4 \lambda(p_h^2 - 1) + 4\Lambda\right]\bigg\},\nonumber\\
C^h_2 & = & \frac{1}{48C_h^3 e^2}\bigg\{2\kappa^{2}(w_h^2-1)^2 +
  4e^2C_h^2(\kappa^{2}w_h^2\eta^2p_h^2 - 1) +
  e^2C_h^4(\kappa^2 (\eta^4 \lambda(p_h^2 - 1)^2 + 4\Lambda)\nonumber\\
&& + \frac{1}{4e^2C_h^4}\bigg[10\kappa^2(w_h^2 - 1)^2 +
  8e^2C_h^2(\eta^2 p_h^2 w_h^2 \kappa^2 - 1) -
  e^2 C_h^4 \kappa^2 (\eta^4 \lambda( p_h^2 - 1)^2 + 4\Lambda))\bigg]\bigg\}\nonumber~.
\end{eqnarray}
 
Much like the near-core expansion, 
the general form near the cosmological horizon is written in 
terms of three constants, in this case $C_h$, $p_h$, and $w_h$.

\section{Time-dependent Equations}
\label{appendix-timelike-eqs}
Following the ansatz of Eq.~(\ref{timelikeeq}), the time-dependent Einstein equations are
\begin{eqnarray}
G^{0}_{0} &=& \frac{3}{B(t)^{2}} - \frac{1}{C(t)^{2}} - 3\left(\frac{B'(t)}{B(t)}\right)^{2}
- 6\frac{B'(t)C'(t)}{B(t)C(t)} -\left(\frac{C'(t)}{C(t)}\right)^{2} 
= \kappa^2 T^{0}_{0} ,\nonumber\\
G^{r}_{r} &=& \frac{1}{B(t)^{2}} - \frac{1}{C(t)^{2}} -\left(\frac{B'(t)}{B(t)}\right)^{2}
-4\frac{B'(t)C'(t)}{B(t)C(t)} - \left(\frac{C'(t)}{C(t)}\right)^{2} -
2\frac{B''(t)}{B(t)} - 2\frac{C''(t)}{C(t)} = \kappa^2 T^{r}_{r} ,\nonumber\\
G^{\theta}_{\theta} &=& \frac{3}{B(t)^{2}} - 3\left(\frac{B'(t)}{B(t)}\right)^{2}
- 3\frac{B'(t)C'(t)}{B(t)C(t)}-3\frac{B''(t)}{B(t)} -
\frac{C''(t)}{C(t)}  = \kappa^2 T^{\theta}_{\theta}~,
\end{eqnarray}
where energy-momentum tensor on the right hand side of these equations
is given by
\begin{eqnarray}
T^{0}_{0} &=& - \bigg[\eta^{2}\left(\frac{p'(t)^{2}}{2} + \frac{p(t)^{2}w(t)^{2}}{C(t)^{2}}\right)
+ \frac{1}{e^{2}C(t)^{2}}\left(w'(t)^{2}+\frac{(1-w(t)^{2})^{2}}{2C(t)^{2}}\right) + 
\frac{\lambda \eta^{4}}{4}(p(t)^{2}-1)^{2}+\Lambda\bigg],\nonumber\\
T^{r}_{r} &=& \eta^{2}\left(\frac{p'(t)^{2}}{2} - \frac{p(t)^{2}w(t)^{2}}{C(t)^{2}}\right)
+ \frac{1}{e^{2}C(t)^{2}}\left(w'(t)^{2} - \frac{(1-w(t)^{2})^{2}}{2C(t)^{2}}\right) 
- \frac{\lambda \eta^{4}}{4}(p(t)^{2}-1)^{2}-\Lambda,\nonumber\\
T^{\theta}_{\theta} &=&  \eta^{2}\frac{p'(t)^{2}}{2} + \frac{(1-w(t)^{2})^{2}}{2 e^{2}C(t)^{4}}
- \frac{\lambda \eta^{4}}{4}(p(t)^{2}-1)^{2}-\Lambda~.
\end{eqnarray}
\\
Eqs.(\ref{eq:higgs}-\ref{eq:yangmills}) become
\begin{equation}
p''(t)+\left(3\frac{B'(t)}{B(t)}+2\frac{C'(t)}{C(t)}\right)p'(t) 
+ \frac{2w(t)^{2}p(t)}{C(t)^{2}}+\lambda\eta^{2}p(t)(p(t)^{2}-1)=0
\end{equation}
and
\begin{equation}
w''(t)+3\frac{B'(t)}{B(t)}w'(t)-\frac{w(t) (1-w(t)^{2})}{C(t)^{2}}+e^{2}\eta^{2}p(t)^{2}w(t)=0~.
\end{equation}

\section{Smooth Brane solution in $dS_6$}
\label{appendix-brane-dS6}

As we explained in the main body of the text, our metric and
matter ansatz allows for the description of solutions that should probably not be considered 
bubbles of nothing in the sense that we use it in the present
context.  They are nevertheless worth mentioning, since they
are interesting geometries in their own right. Perhaps the most important example of this is the
type of solutions which describe
the nucleation of smooth magnetically charged de Sitter branes in $dS_6$. Similar
instanton solutions have been recently discussed in the literature 
\cite{BPSPV-2} using black branes. The
difference between these two type of instantons is the existence
of a horizon on the black brane solutions that is not present 
in the smooth cores that we study here. The 
possibility of a smooth solitonic brane is again due to the
fact that we have extended our model to include additional degrees
of freedom that resolve the singularities of the Dirac monopole
solution. 

\begin{figure}[htbp]
\centering\leavevmode
\epsfysize=8cm \epsfbox{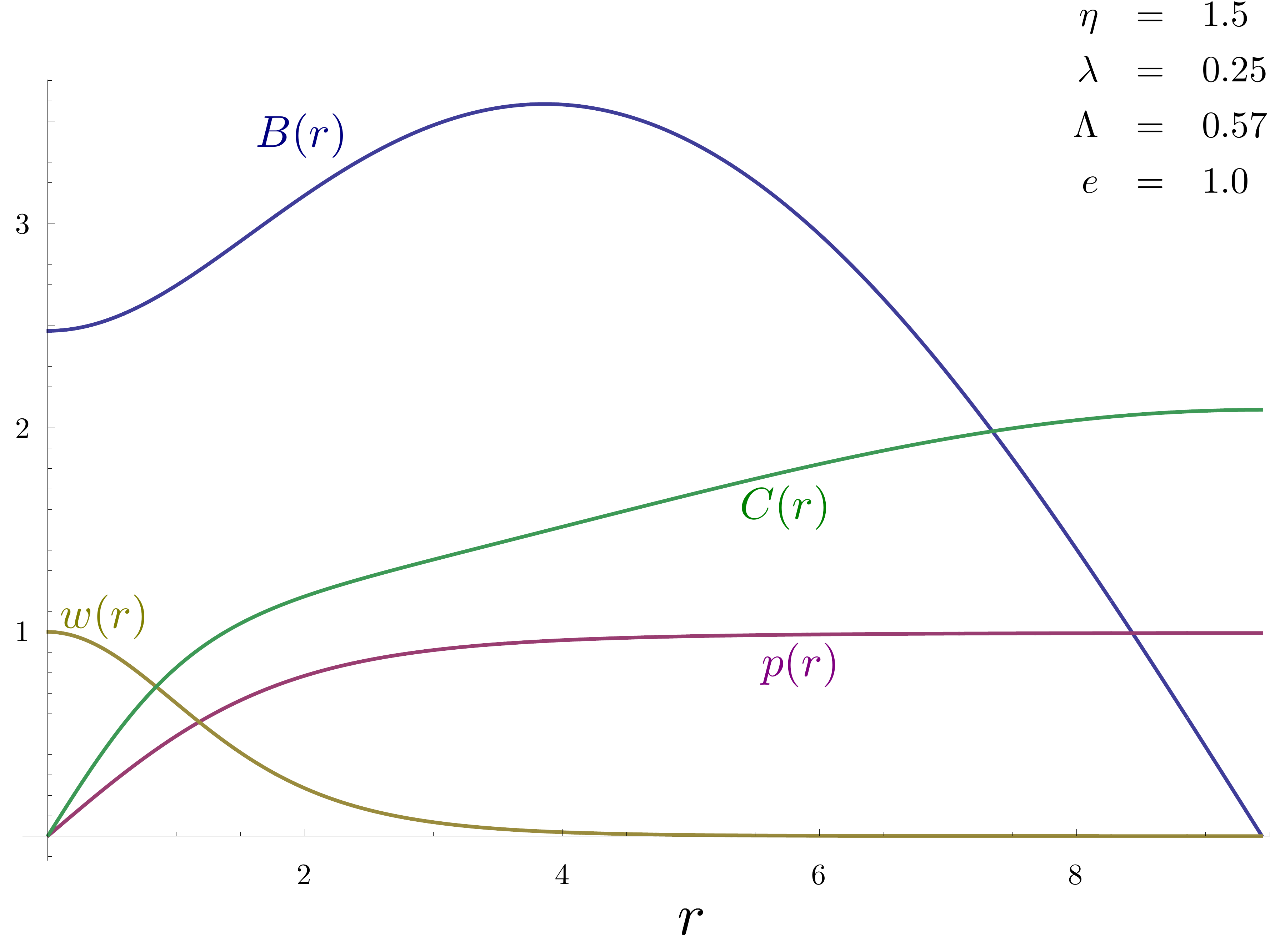}
\caption {The near-core region of what appears to be a bubble of nothing, but which does not in fact asymptote to a $4d$ region.}
\label{braned6-space}
\end{figure}

\begin{figure}[htbp]
\centering\leavevmode
\epsfysize=12cm \epsfbox{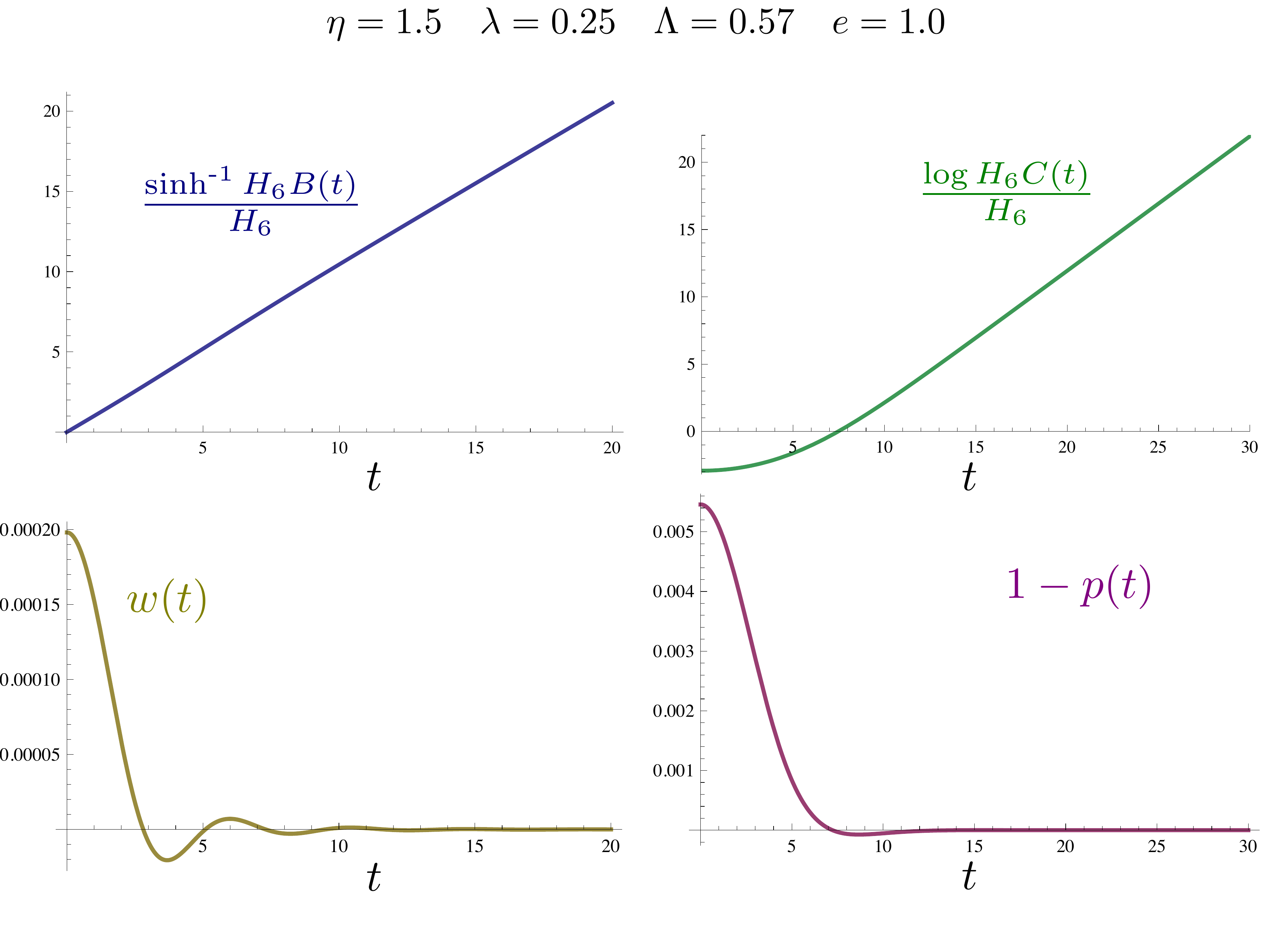}
\caption {The decompactification of the $4d$ geometry surrounding the faux bubble of nothing geometry found in Fig.~(\ref{braned6-space}).}
\label{braned6-time}
\end{figure}

By choosing appropriate values for the parameters in the theory
one can find solutions of this type. We show in
Fig.~(\ref{braned6-space}) the instanton solution within
region I of the spacetime [see  Fig.~(\ref{confdiagdSBON})].
The solution in this region is rather similar to the bubble of
nothing solutions obtained in the main body of the text. However, once we continue across the lightcone into region II, things 
fall apart. [see  Fig.~(\ref{braned6-time})].
In particular the ``radion" $C(t)$ grows without bound, signaling
the decompactification of spacetime. This shows that 
one cannot consider this solution to be relevant
for a compactified spacetime since its global structure is markedly different from anything four dimensional. 
On the other hand, we can see that
the form of the metric rapidly approaches that of Eq.~(\ref{aniso-dS6}), the anisotropic 
ansatz for $dS_6$.  This validates the interpretation of this
instanton as the quantum mechanical creation of a $2+1$ dimensional solitonic de Sitter
brane in a $dS_6$ bulk.

\section{Numerical Techniques}
\label{appendix-multiple-shooting}
The defining field configuration of a bubble of nothing is the existence
of a smooth core where the extradimensional fiber degenerates.  In the
case at hand, this requirement leaves three unknown initial field
values for a seventh order ordinary differential system.  These
three parameters must be determined by evolving the fields between the
core and a sufficiently asymptotic region, where the boundary conditions are known.

The three unknown parameters can be obtained by treating them as
``shooting parameters." This means a guess is made, followed by the
numerical evolution of the resulting solution toward the asymptotic boundary
conditions.  If the numerical evolution fails to asymptote to the
boundary conditions, the shooting parameters are appropriately
adjusted, and the procedure is repeated.  However this method is
rendered intractable by the exponentially growing modes of the differential equations.
Because we would like to evolve the fields across distances much
longer than the Compton wavelength of the most massive of the four
fields, the behavior is extremely sensitive to the initial
conditions. This would require one to maintain tremendous numerical
precision throughout the evolution of the solution.  The inefficiency
arises because one must cancel the coefficient of the three growing modes
to far more numerical precision than the eventual solution warrants.

A solution to this problem is easily implemented using the so-called
multiple shooting method \cite{shooting-method} .  The integration interval is
divided into many subintervals of length less than the Compton
wavelengths of the fields.  Since each interval is small, the
numerical evolution depends approximately linearly on the initial conditions. The
many intervals are then pieced together by demanding that each
function be continuous and differentiable across the boundaries of the
subintervals. This is achieved using Newton's method.  By
extrapolating the mismatch between all neighboring subintervals as a
function of the shooting parameters and multiple boundary conditions,
an optimal guess for the improvement of the shooting parameters can be
made. In this sense, the solution is found by a combination of
shooting and relaxing of the fields between subintervals.  A
non-linear problem in few dimensions is traded for a linear problem in
many dimensions.  The approximately linear nature of the problem is
important for Newton's method to work.

\end{document}